\documentclass[10pt]{article}

\newcommand{\papertitle}{Can Foundation Models Hear What Made That Sound?\\\large A Tiered Benchmark of Audio-Language Models and Traditional Classifiers for Closed-Set Sound Source Identification}
\newcommand{\shorttitle}{Sound Source Identification Benchmark}
\newcommand{\organizationname}{Perle}
\newcommand{\paperdate}{\today}

\usepackage{iftex}
\ifPDFTeX
  \usepackage[utf8]{inputenc}
  \usepackage[T1]{fontenc}
  \usepackage{lmodern}
\else
  \usepackage{fontspec}
\fi

\usepackage{microtype}

\usepackage[
  top=2.5cm,
  bottom=2.8cm,
  left=2.5cm,
  right=2.5cm
]{geometry}

\usepackage{amsmath}
\usepackage{amssymb}
\usepackage{graphicx}
\usepackage{float}
\usepackage[
  labelfont=bf,
  font=small,
  justification=justified,
  singlelinecheck=false
]{caption}
\usepackage{booktabs}
\usepackage{array}
\usepackage{ragged2e}
\usepackage[table]{xcolor}
\usepackage{tabularx}
\usepackage{enumitem}
\usepackage{fancyhdr}
\usepackage{titlesec}
\usepackage{longtable}
\usepackage{multirow}
\usepackage{xurl}
\usepackage[
  colorlinks=true,
  linkcolor=black,
  citecolor=black,
  urlcolor=black
]{hyperref}

\newcolumntype{C}{>{\centering\arraybackslash}X}
\newcolumntype{L}{>{\raggedright\arraybackslash}X}
\newcolumntype{P}[1]{>{\RaggedRight\arraybackslash}p{#1}}
\DeclareRobustCommand{\modelname}[1]{%
  \texttt{\hyphenchar\font=45\relax #1}}

\definecolor{brandcolor}{HTML}{000000}
\definecolor{lightlinecolor}{HTML}{CCCCCC}
\definecolor{rowgray}{HTML}{F5F5F5}

\titleformat{\section}
  {\normalfont\large\bfseries}{\thesection.}{0.5em}{}
\titlespacing{\section}{0pt}{18pt plus 3pt minus 2pt}{6pt}

\titleformat{\subsection}
  {\normalfont\normalsize\bfseries}{\thesubsection.}{0.5em}{}
\titlespacing{\subsection}{0pt}{12pt plus 2pt minus 1pt}{4pt}

\titleformat{\subsubsection}[runin]
  {\normalfont\normalsize\bfseries}{\thesubsubsection.}{0.5em}{}[.\quad]
\titlespacing{\subsubsection}{0pt}{8pt plus 1pt}{0pt}

\setlist[itemize]{leftmargin=1.5em, topsep=3pt, itemsep=2pt, parsep=0pt}
\setlist[enumerate]{leftmargin=1.5em, topsep=3pt, itemsep=2pt, parsep=0pt}

\fancypagestyle{firstpage}{
  \fancyhf{}
  
  \fancyfoot[L]{\footnotesize
    \textcopyright~\the\year~\organizationname. All rights reserved.}
  \fancyfoot[R]{\small 1}
}

\renewenvironment{abstract}{%
  \small\bfseries\noindent\ignorespaces
}{\par\medskip}

\newcommand{\brandrule}{%
  \noindent\textcolor{brandcolor}{\rule{\linewidth}{1.5pt}}\par}
\newcommand{\lightrule}{%
  \noindent\textcolor{lightlinecolor}{\rule{\linewidth}{0.4pt}}\par}
\newcommand{\emailaddr}[1]{\href{mailto:#1}{\nolinkurl{#1}}}

\begin{document}
\thispagestyle{firstpage}

\begin{flushleft}
  \IfFileExists{company_logo.png}{%
    \includegraphics[height=20pt]{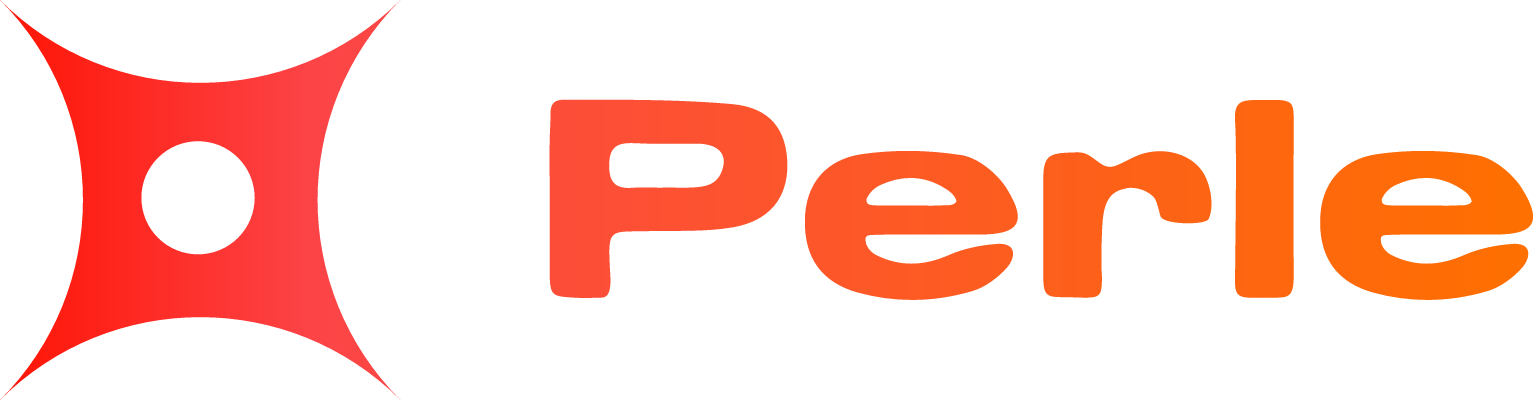}%
  }{%
    \textbf{\organizationname}%
  }
  \hfill
  {\small\paperdate}
\end{flushleft}

\vspace{4pt}
\brandrule
\vspace{8pt}

\noindent{\fontsize{17pt}{20pt}\selectfont\bfseries \papertitle\par}

\vspace{10pt}

\noindent
\textbf{Sajjad Abdoli}\textsuperscript{1,*,\dag}\footnote{Corresponding author: \emailaddr{sajjad@perle.ai}},\allowbreak\quad
\textbf{Ghassan Al-Sumaidaee}\textsuperscript{1,*,\dag}\footnote{Corresponding author: \emailaddr{ghassan.al-sumaidaee@perle.ai}; ORCID: \href{https://orcid.org/0000-0002-5536-0252}{0000-0002-5536-0252}},\allowbreak\quad
\textbf{Ahmad ElShiekh}\textsuperscript{1,*}\footnote{ORCID: \href{https://orcid.org/0009-0001-6837-6202}{0009-0001-6837-6202}; current contact: \emailaddr{ahmadelshiekh9@gmail.com}},\allowbreak\quad
\textbf{Ahmed Rashad}\textsuperscript{1}
\par\vspace{3pt}
{\small\textsuperscript{1}Perle\quad
\textsuperscript{*}Equal contribution; names sorted alphabetically.\quad
\textsuperscript{\dag}Corresponding authors\\[2pt]
\emailaddr{sajjad@perle.ai}\quad
\emailaddr{ghassan.al-sumaidaee@perle.ai}\\[2pt]
\emailaddr{mad.elshiekh@perle.ai}\quad
\emailaddr{ahmed@perle.ai}}

\vspace{10pt}

\begin{abstract}
We benchmark eleven audio classification methods — five task-aware closed-set LLMs
(four Gemini models plus open-weight Kimi-Audio-7B-Instruct), four fixed-vocabulary
taggers (YAMNet, PANNs, Whisper-AT, SSLAM), a zero-shot audio-text model (CLAP), and
an audio-grounded LLM (BAT) — on a closed-set sound-source identification task over
2{,}242 clips spanning 23 fine-grained classes (11 categories). Since these methods
differ fundamentally in how they receive the task and how outputs are scored, we
group them into four evaluation tiers rather than one leaderboard, reporting macro
Precision, Recall, F1, and false-negative rate per tier. The best model
(Gemini-3.1-Pro-Preview) reaches 85.6\% category-level and 56.7\% fine-grained F1;
Kimi-Audio is competitive for its size (67.5\%/32.9\%) but fails to answer 1.6\% of
samples. SSLAM and CLAP match or exceed the best closed-set model at
the category level without seeing the candidate list, but fall behind at the
fine-grained level. Analyzing the Gemini models' chain-of-thought across 8{,}968
responses, we find response length does not predict accuracy, an apparent
"holistic judgment beats detailed analysis" effect is better explained as a
difficulty confound, and wrong answers are stated confidently 92--100\% of the
time. We report full per-class confusion matrices and metrics for all eleven
methods, identify the structural error modes behind most of the accuracy loss
between granularities, and give practical guidance for choosing among these method
families.
\end{abstract}


\vspace{6pt}
\lightrule
\vspace{14pt}

\section{Introduction}
\label{sec:introduction}

Identifying the source of an everyday sound — a doorbell, a siren, a running tap —
is a task humans perform almost effortlessly, but it has historically required
purpose-built audio classifiers \cite{ABDOLI2019252} trained on fixed label vocabularies drawn from
ontologies such as AudioSet~\cite{audioset2017}. The recent emergence of
general-purpose multimodal large language models (LLMs) with native audio input,
alongside a new generation of audio-grounded LLMs~\cite{zheng2024bat,gong2024ltu} and
self-supervised audio encoders~\cite{sslam2025}, raises a practical question for
anyone building a product on top of audio understanding: given a fixed set of sounds
we care about, which class of method should we use, and how much can we trust its
answer?

This is harder to answer cleanly than it first appears, because these method
families are not interchangeable in how they are evaluated. A multimodal LLM can be
handed an explicit list of candidate answers and asked to pick one; a fixed-vocabulary
audio tagger has no notion of a candidate list at all and must be scored by mapping
its native output vocabulary onto the target taxonomy after the fact; a zero-shot
audio-text similarity model is given natural-language descriptions of the candidates
and ranks them by embedding similarity; and an audio-grounded LLM fine-tuned
primarily for open-ended dialogue may produce free text that has to be graded by a
second model. Comparing raw accuracy numbers across these four setups risks
conflating "this method understood the sound better" with "this method's task was
easier."

We address this by running all eleven methods on the identical 2{,}242-sample,
23-class balanced dataset, and by treating the four evaluation paradigms as distinct
tiers rather than collapsing them into one leaderboard. We report standard
per-class detection metrics (macro Precision, Recall/TPR, F1, FNR) rather than raw
accuracy, since these are the conventional reporting choice in the environmental
sound classification (ESC) literature~\cite{piczak2015esc50} and remain well-defined
even for the one method (BAT) for which a full multiclass confusion matrix cannot be
constructed. We further exploit the fact that the four LLMs in our study produce
free-text chain-of-thought reasoning, and run a secondary LLM-based classification
pass over all 8{,}968 saved responses to characterize \emph{how} these models reason
their way to an answer — not just whether they are right.

The main contributions of this paper are:

\begin{itemize}
  \item A four-tier evaluation framework for comparing task-aware closed-set
        selection, task-agnostic fixed-vocabulary tagging, zero-shot audio-text
        similarity, and open-vocabulary LLM-judged scoring on the same underlying
        data, with an explicit accounting of why cross-tier comparisons must be
        read directionally rather than as a strict ranking.
  \item A full benchmark of eleven methods across four families on a 2{,}242-sample,
        23-class (11-category) closed-set sound-source identification task, with
        complete per-class Precision/Recall/F1/TPR/FNR metrics and confusion
        matrices for every method at both granularities.
  \item An analysis of 8{,}968 LLM chain-of-thought traces showing that (a) response
        length does not predict fine-grained accuracy, (b) an apparent
        strategy-accuracy correlation is confounded by which classes elicit which
        strategy, and (c) none of the four LLMs reliably signals uncertainty when
        wrong, with 92--100\% of incorrect answers phrased with confident language.
\end{itemize}

\section{Background and Related Work}
\label{sec:background}

Environmental sound classification (ESC) is typically framed as a fixed-vocabulary
multiclass or multilabel problem: a model is trained (or evaluated) against a
closed ontology such as the 527-class AudioSet ontology~\cite{audioset2017} or the
50-class ESC-50 dataset~\cite{piczak2015esc50}. Classical approaches to this problem
are convolutional or transformer audio taggers trained end-to-end on labeled audio,
of which YAMNet~\cite{yamnet,hershey2017cnn} and PANNs~\cite{kong2020panns} are
widely used pretrained baselines. More recent work adapts speech-recognition
backbones for tagging (Whisper-AT~\cite{gong2023whisperat}) or introduces
self-supervised pretraining objectives specifically designed for overlapping,
polyphonic soundscapes (SSLAM~\cite{sslam2025}).

A parallel line of work moves away from fixed vocabularies entirely.
Contrastive audio-text models such as CLAP~\cite{elizalde2023clap} embed audio and
natural-language descriptions into a shared space, enabling zero-shot classification
against arbitrary text prompts rather than a fixed label set. Most recently,
general-purpose multimodal LLMs with native audio understanding (e.g., the Gemini
family~\cite{gemini2025}) and audio-grounded instruction-tuned LLMs such as
LTU~\cite{gong2024ltu}, BAT~\cite{zheng2024bat}, and Kimi-Audio~\cite{kimiaudio2025}
reframe audio classification as a language generation or selection task, in
principle allowing free-form reasoning before an answer is produced. Kimi-Audio
in particular is an open-weight 7B audio-native LLM (an audio tokenizer, a
Qwen2.5-based transformer core, and an audio detokenizer for optional speech
output), distinguishing it architecturally from both the much larger proprietary
Gemini models and BAT's LoRA-adapted encoder-plus-frozen-LLM design.

All three of Kimi-Audio, BAT, and (almost certainly) Gemini share a common
architectural commitment despite their differences: a separately pretrained audio
or speech \emph{encoder} converts the raw signal into a compressed representation
before the LLM ever sees it. Concurrent work by
Fan~et~al.~\cite{fan2026llmreadspectrogramencoderfree} (Mel-LLM) questions whether this encoder is
necessary at all, removing it entirely and projecting lightly-processed
Mel-spectrogram patches directly into the LLM via a single linear layer — the
speech-domain analogue of encoder-free vision-language modeling. Their central
finding is a trade-off directly relevant to
sound-source identification: the encoder-free model shows the largest gains on
tasks where the answer is carried by low-level acoustic structure rather than
lexical content (including a 50-class environmental sound classification
benchmark, ESC-50~\cite{piczak2015esc50}, where it improves substantially over an
otherwise-comparable encoder-based baseline), while remaining weaker on
knowledge-intensive spoken question answering. This suggests that how much an
audio-LLM's front end compresses toward semantic abstraction, versus preserving
raw acoustic detail, is itself a variable worth manipulating for tasks like ours —
a possibility we return to in Section~\ref{sec:future-work}.

These four families — fixed-vocabulary taggers, zero-shot audio-text models,
general-purpose multimodal LLMs, and audio-grounded instruction-tuned LLMs — differ
in a way that matters for benchmarking: a fixed-vocabulary tagger has no mechanism
for accepting an externally-defined candidate list at inference time, while an LLM
has no native scoring mechanism over a fixed label set without first being given
one as part of the prompt. Comparing across families therefore risks conflating a
model's audio understanding with how much task information it was given. This is
why our study deliberately keeps the underlying audio, ground truth, and taxonomy
identical across all eleven methods (Section~\ref{sec:methodology}), but treats the
resulting comparison as tiered by task-awareness and scoring mechanism rather than
flat, making the source of any given accuracy gap between methods (audio
understanding versus task framing) explicit rather than implicit.

\subsection{Other General-Purpose Multimodal LLMs Considered}
\label{sec:other-llms}

Beyond the four Gemini models and Kimi-Audio, Tier A of this study
(Section~\ref{sec:methodology}) excludes several other general-purpose multimodal
LLMs we investigated, because we could not confirm they support native audio input
at the time of this study:

\begin{itemize}
  \item \textbf{Anthropic's Claude models} do not accept audio as an input
        modality at all (text and image only, per Anthropic's own model
        overview\footnote{\url{https://platform.claude.com/docs/en/about-claude/models/overview}} — ``All current Claude models support text and image input''),
        so no Claude model could be included in Tier A.
  \item \textbf{OpenAI's frontier reasoning models} (the GPT-5.x family) likewise do
        not accept audio input: OpenAI's own model documentation lists audio as
        ``Not supported'' for GPT-5\footnote{\url{https://developers.openai.com/api/docs/models/gpt-5}}.
        OpenAI's dedicated audio models (\texttt{gpt-audio}, \texttt{gpt-audio-mini})
        do accept audio, per OpenAI's audio-model
        guide\footnote{\url{https://developers.openai.com/api/docs/guides/audio}},
        and were smoke-tested on a small sample of this benchmark, but were not
        scaled to the full 2{,}242-sample run reported here and are therefore
        omitted from Tier A.
  \item \textbf{Meta's Muse Spark}, released via the Meta Model API, is
        advertised as a multimodal reasoning model, but Meta's own developer
        documentation\footnote{\url{https://ai.developer.meta.com/docs/getting-started/models}}
        lists only text, image, video, and PDF as supported input modalities —
        audio is not listed. We tested this directly by attempting to upload one
        of our benchmark audio clips to Meta's consumer-facing assistant at
        \texttt{meta.ai}, which rejected the file outright with ``Unsupported
        file format. Please upload an image.'' With the primary documentation
        not listing audio support and a direct upload attempt rejecting audio
        files entirely, we did not include Muse Spark in this study.
  \item \textbf{Moonshot AI's Kimi K3} (the general-purpose chat model,
        distinct from Kimi-Audio) likewise does not accept audio: its official
        documentation overview\footnote{\url{https://platform.kimi.ai/docs/overview}}
        and K3 quickstart
        guide\footnote{\url{https://platform.kimi.ai/docs/guide/kimi-k3-quickstart}}
        list only text, image, and video as supported inputs, with no audio
        modality documented anywhere in the platform's indexed documentation. We
        note this explicitly because both models share a developer and a name
        prefix; only \textbf{Kimi-Audio}, a separate, dedicated open-weight audio
        foundation model~\cite{kimiaudio2025}, was included in this study
        (Tier A).
\end{itemize}

We flag this explicitly because it is a real constraint on how far the Tier A
comparison generalizes: it spans four Gemini model versions and one open-weight
audio-native model, not the full landscape of general-purpose multimodal LLMs, and
should be read accordingly.

\section{Data}
\label{sec:data}

\subsection{Collection and Class Structure}

The benchmark dataset consists of 2{,}242 short audio clips drawn from a curated,
balanced subset of everyday sound recordings. Each clip is labeled with a
\texttt{sound\_type} (broad category) and a \texttt{sub\_category} (fine-grained
sub-type); the unique combination of the two fields defines 23 fine-grained classes,
which collapse into 11 broader categories. Table~\ref{tab:dataset-summary} lists
both levels; the full 23-class list with its lettered candidate mapping is given in
Appendix~\ref{app:candidates}.

\subsection{Dataset Summary}

The dataset is balanced at approximately 100 samples per fine-grained class, with one
exception (\texttt{Siren - Other}, 42 samples) reflecting natural data availability
rather than a sampling choice.

\begin{table}[H]
\centering
\caption{Dataset structure. All 2{,}242 samples were attempted by all eleven methods.}
\label{tab:dataset-summary}
\small
\renewcommand{\arraystretch}{1.25}
\begin{tabularx}{\textwidth}{L C}
  \toprule
  \textbf{Property} & \textbf{Value} \\
  \midrule
  \rowcolor{rowgray}
  Total samples & 2{,}242 \\
  Category-level classes (\texttt{sound\_type}) & 11 \\
  \rowcolor{rowgray}
  Fine-grained classes (\texttt{sound\_type} + \texttt{sub\_category}) & 23 \\
  Samples per fine-grained class & $\sim$100 (42--104) \\
  \rowcolor{rowgray}
  Methods evaluated & 11 \\
  Samples evaluated per method & 2{,}242 (100\% coverage; 0 unresolved errors for ten
  methods, 36/2{,}242 unresolved for Kimi-Audio — Section~\ref{sec:methodology}) \\
  \bottomrule
\end{tabularx}
\end{table}

\section{Methodology}
\label{sec:methodology}

\subsection{Study Design}

Every method was evaluated on the identical 2{,}242-sample set and scored against the
identical ground truth at both the category and fine-grained level. What differs
across methods is \emph{how} each one is given the task and how its raw output is
converted into a predicted class. We group the eleven methods into four evaluation
tiers along these two axes.

\subsubsection{Tier A --- Task-aware closed-set, exact match}

Four Gemini models (\modelname{gemini-2.5-flash}, \modelname{gemini-2.5-pro},
\modelname{gemini-3.5-flash}, \modelname{gemini-3.1-pro-preview}) were each given the audio
clip together with a fixed, lettered list of all 23 candidate labels (e.g.,
\texttt{A. Appliance Alarm - Appliance Alarm}, \ldots, \texttt{W. Yell - Yell}), an
explicit instruction to reason freely (chain-of-thought), and a requirement to commit
to exactly one letter as the final answer (full prompt text in
Appendix~\ref{app:prompt-tiera}). Scoring is exact string match between the
predicted and true label at each granularity.

We additionally tested \textbf{Kimi-Audio-7B-Instruct}~\cite{kimiaudio2025}, an
open-weight audio-native LLM, with the identical prompt format. Unlike
BAT (Tier D), Kimi-Audio is instruction-tuned broadly enough to follow the
closed-set letter-selection task, making it a genuine Tier A member rather than a
forced fallback to open-vocabulary scoring. In practice its compliance with the
"share your reasoning" instruction was inconsistent: the large majority of
responses were a bare letter with no reasoning text at all, a minority included full
chain-of-thought reasoning ending in the requested JSON answer, and 36/2{,}242
samples (1.6\%) produced no scoreable answer whatsoever — 32 repetition-loop
generations that never concluded (the model restates a description in a loop until
hitting the generation-length limit) and 4 explicit refusals (e.g., ``none of the
candidates match this sound''; one instance, ``I cannot analyze the sound as there
is no audio provided,'' hallucinates the \emph{absence} of input that was in fact
given — the mirror image of the Gemini grounding failure in
Section~\ref{sec:reasoning} that hallucinates the \emph{presence} of input never
given, and then answers anyway rather than abstaining). These 36 samples are
excluded from Kimi-Audio's reported metrics at both granularities, identically to
how an audio-file error would be excluded for any other method; unlike BAT's
exclusions, these are not a scoring
design choice but genuine non-answers we could not parse a letter from even after
two rounds of extraction-logic refinement (JSON, LaTeX-boxed, bare-letter,
natural-language, and verbatim-candidate-echo matching) and a same-settings retry
that reproduced the identical failures, confirming they are not transient.

\subsubsection{Tier B --- Task-agnostic fixed-vocabulary, post-hoc mapped}

Four models (YAMNet~\cite{yamnet}, PANNs~\cite{kong2020panns},
Whisper-AT~\cite{gong2023whisperat}, SSLAM~\cite{sslam2025}) were run with no
knowledge of our 23-class taxonomy at all. Each produces a ranked list of scores over
its native $\sim$521--527-class AudioSet vocabulary. We restrict this ranking to the
subset of AudioSet labels with a known mapping into our taxonomy (tagged \emph{exact}
or \emph{approximate} match quality by hand) and take the top-ranked mapped label as
the prediction. Because these models never see the candidate set, their effective
task is harder — ranking over the full AudioSet ontology — than the 23-way selection
given to Tier A.

\subsubsection{Tier C --- Zero-shot audio-text similarity}

Microsoft CLAP~\cite{elizalde2023clap} is given the same 23 candidates as Tier A, but
phrased as natural-language descriptions (e.g., \emph{"a police car siren"}) rather
than lettered options (full candidate phrasings in Appendix~\ref{app:prompt-tierc}),
and produces a prediction via cosine similarity between the
audio embedding and each candidate's text embedding, without any generative reasoning
step. CLAP is never separately queried against an 11-candidate category-level set:
as with Tier A, both granularities are scored from this single 23-way ranking —
Fine-Grained accuracy compares the full top-ranked candidate against the true
label, and Category-Level accuracy compares only its \texttt{sound\_type} component
against the true category, rather than from an independently-elicited coarser
prediction.

\subsubsection{Tier D --- Open-vocabulary generation with LLM-judged scoring}

BAT~\cite{zheng2024bat} (a Spatial-AST audio encoder coupled with a LoRA-adapted
LLaMA-2-7B via the SLAM-LLM framework) was run in its native open-vocabulary
free-text question-answering mode (exact question text in
Appendix~\ref{app:prompt-tierd}), since a direct experiment giving BAT the same
lettered candidate list as Tier A failed: BAT's LoRA fine-tuning produces a strong
bias toward binary yes/no answers regardless of instruction, and could not be
induced to select a letter. BAT's free-text answers were instead scored by an
independent LLM judge (\modelname{gemini-2.5-flash-lite}), prompted to determine
separately whether the answer correctly identifies the category-level and
fine-grained sound source, allowing for synonyms and paraphrases (full judge prompt
in Appendix~\ref{app:prompt-judge}).

\subsection{Evaluation Metrics}

For every method and every sample, we treat each of the $C$ classes ($C=11$ at the
category level, $C=23$ at the fine-grained level) as a one-vs-rest detection problem
and compute, per class $c$: true positives $\mathrm{TP}_c$, false positives
$\mathrm{FP}_c$, false negatives $\mathrm{FN}_c$, and true negatives
$\mathrm{TN}_c$. From these we derive

\begin{equation}
  \mathrm{Precision}_c = \frac{\mathrm{TP}_c}{\mathrm{TP}_c + \mathrm{FP}_c}, \qquad
  \mathrm{Recall}_c \;(\mathrm{TPR}_c) = \frac{\mathrm{TP}_c}{\mathrm{TP}_c + \mathrm{FN}_c},
  \label{eq:pr}
\end{equation}

\begin{equation}
  F1_c = \frac{2 \cdot \mathrm{Precision}_c \cdot \mathrm{Recall}_c}{\mathrm{Precision}_c + \mathrm{Recall}_c}, \qquad
  \mathrm{FNR}_c = 1 - \mathrm{Recall}_c = \frac{\mathrm{FN}_c}{\mathrm{TP}_c + \mathrm{FN}_c}.
  \label{eq:f1fnr}
\end{equation}

We report the unweighted (macro) average of each quantity across all $C$ classes as
the headline number for a method at a given granularity. We use macro-averaged F1,
TPR, and FNR rather than raw accuracy as our primary reported metrics, for two
reasons. First, this is the conventional reporting choice in the ESC
literature~\cite{piczak2015esc50,kong2020panns}, where per-class detection rate and
false-negative rate are more informative than a single pooled accuracy figure when
class difficulty is uneven (as it clearly is here — see
Section~\ref{sec:results}). Second, macro Recall/TPR and FNR remain well-defined for
Tier D (BAT), where no discrete predicted class exists (the output is free text
judged correct/incorrect per sample), so no full confusion matrix — and therefore no
Precision or F1 — can be constructed; reporting only what is actually measurable for
each tier, rather than forcing every method into the same metric, is itself part of
the tiered-comparison discipline in Section~\ref{sec:results}. For completeness and
comparability with the accompanying project write-up, raw accuracy for every method
is also given in Appendix~\ref{app:full-results}.

\subsubsection{Match-quality tagging for Tier B}
\label{sec:match-quality}

Tier B methods never see our 23-class taxonomy; each one tags audio using its own
native AudioSet vocabulary ($\sim$521--527 classes), and a hand-built lookup table translates the
relevant AudioSet labels into our \texttt{(sound\_type,\allowbreak\ sub\_category)} pairs. Each
entry in that lookup table is tagged \emph{exact} or \emph{approximate}, describing
the quality of the \emph{translation itself} — not the model's confidence, and not
how close a wrong guess was.

\begin{itemize}
  \item \textbf{\emph{exact}}: AudioSet has a label that specifically corresponds to
        our sub-category. For example, the true label \texttt{Siren - Police}
        matched against YAMNet's raw tag \texttt{"Police car (siren)"}, which the
        lookup table maps to \texttt{(Siren, Police, exact)}. AudioSet has a
        genuinely specific concept here, so a hit reflects the model actually
        detecting that particular sound.
  \item \textbf{\emph{approximate}}: AudioSet has no label specific enough to
        express our sub-category, only a broader, generic one, and we chose to map
        that generic bucket onto the closest available sub-category when building
        the lookup table. For example, the true label
        \texttt{Appliance Alarm - Appliance Alarm} matched against YAMNet's raw tag
        \texttt{"Alarm"}, which the lookup table maps to
        \texttt{(Appliance Alarm, Appliance Alarm, approximate)}. AudioSet's
        \texttt{"Alarm"} label is a single bucket that also covers car alarms,
        smoke alarms, and alarm clocks; the model is not telling us ``this is
        specifically an appliance,'' only ``this sounds like some kind of alarm,''
        and our table is the one deciding that generic signal counts as a match for
        this particular sub-category.
\end{itemize}

Critically, this tag has \emph{no effect on scoring}. Whether a mapped prediction is
\emph{exact} or \emph{approximate}, it is scored identically to every other
prediction in this study: does the translated label string equal the true label
string? A match is full credit regardless of tag; a non-match is a miss regardless
of tag — there is no partial credit for \emph{approximate} matches and no penalty
either. \texttt{match\_quality} is recorded purely as a diagnostic column for
post-hoc error analysis (e.g., quantifying how many of a method's correct answers
reflect a genuinely specific detection versus a generic AudioSet bucket defaulting
onto the right sub-category), not as an input to any reported metric. The complete
lookup table, with every \emph{approximate} entry and the reason it could not be
tagged \emph{exact}, is given in Appendix~\ref{app:taxonomy-mapping}.

For four sub-category distinctions (clap count, doorbell type, knock surface,
water-running distance), AudioSet's ontology has no label capable of expressing the
distinction at all, only a category-level one (e.g.\ a generic \texttt{"Doorbell"}
label with no digital/mechanical/wireless variants). When a Tier B method's
top-ranked mapped prediction resolves to one of these category-only entries, it is
scored as an ordinary Fine-Grained miss against whichever true sub-category the
sample actually had — the same treatment any other incorrect prediction receives,
with no special exclusion or partial credit. This keeps Tier B's Fine-Grained
metric on the same footing as Tier A and Tier C rather than granting it credit for
distinctions its underlying vocabulary cannot express.

\subsubsection{Reasoning-strategy analysis (Gemini models only)}

This secondary analysis covers only the four Gemini models, not Kimi-Audio: it
requires a substantial, consistently-produced chain-of-thought to classify, and (as
noted above) the large majority of Kimi-Audio's responses are a bare letter with no
reasoning text at all, making it a poor candidate for this specific pass. We saved
the full reasoning trace for all $2{,}242 \times 4 = 8{,}968$ Gemini responses and
passed each one through a second LLM classification pass
(\modelname{gemini-2.5-flash-lite}) that labeled its primary reasoning strategy
(\emph{detailed acoustic analysis}, \emph{holistic pattern match}, \emph{process of
elimination}, \emph{direct assertion}, or \emph{non-grounded}) and its apparent
confidence level (\emph{high}, \emph{medium}, \emph{low}, or \emph{hedging}), based
purely on the language used — not a self-reported score, since the original prompt
never asked for one. We combined this with simple lexical statistics (word count,
hedge-term and acoustic-term frequency) computed directly from the text.

\section{Results}
\label{sec:results}

\subsection{Category-Level Results (11 classes)}

Table~\ref{tab:results-category} reports macro Precision, Recall (TPR), F1, and FNR
for all eleven methods at the category level, grouped by evaluation tier.

\begin{table}[H]
\centering
\caption{Category-level (11-class) macro-averaged results, by evaluation tier.
$N{=}2{,}242$ for every method except Kimi-Audio ($N{=}2{,}206$; 36 samples produced
no scoreable answer — see Section~\ref{sec:methodology}, marked $\ddagger$). BAT
(Tier D) has no discrete predicted class, so Precision and F1 are not computable
($\dagger$); only Recall/TPR (fraction of each true class the judge scored correct)
and FNR are reported.}
\label{tab:results-category}
\small
\renewcommand{\arraystretch}{1.2}
\begin{tabularx}{\textwidth}{L L C C C C}
  \toprule
  \textbf{Tier} & \textbf{Method} & \textbf{Precision} & \textbf{Recall (TPR)} & \textbf{F1} & \textbf{FNR} \\
  \midrule
  \rowcolor{rowgray}
  & gemini-3.1-pro-preview & 86.4\% & 87.4\% & 85.6\% & 12.6\% \\
  \rowcolor{rowgray} & gemini-2.5-pro         & 79.0\% & 78.3\% & 77.0\% & 21.7\% \\
  \rowcolor{rowgray} & gemini-3.5-flash       & 78.1\% & 78.9\% & 76.7\% & 21.1\% \\
  \rowcolor{rowgray} & kimi-audio-7b-instruct$^\ddagger$ & 71.0\% & 70.3\% & 67.5\% & 29.7\% \\
  \rowcolor{rowgray} \multirow{-5}{*}{A} & gemini-2.5-flash & 69.5\% & 66.3\% & 65.7\% & 33.7\% \\
  \midrule
  \multirow{4}{*}{B} & sslam       & 87.9\% & 87.8\% & 87.3\% & 12.2\% \\
                     & panns       & 84.6\% & 83.9\% & 83.2\% & 16.1\% \\
                     & whisper-at  & 78.5\% & 76.8\% & 76.2\% & 23.2\% \\
                     & yamnet      & 62.2\% & 53.4\% & 52.4\% & 46.6\% \\
  \midrule
  \rowcolor{rowgray}
  C & clap & 89.1\% & 87.3\% & 87.8\% & 12.7\% \\
  \midrule
  D & bat & n/a$^\dagger$ & 57.7\% & n/a$^\dagger$ & 42.3\% \\
  \bottomrule
\end{tabularx}
\end{table}

\subsection{Fine-Grained Results (23 classes)}

Table~\ref{tab:results-finegrained} reports the same metrics at the fine-grained
level, where the model must distinguish the exact sub-category (e.g.,
\texttt{Siren - Police} vs.\ \texttt{Siren - Ambulance}), not just the broad category.

\begin{table}[H]
\centering
\caption{Fine-grained (23-class) macro-averaged results, by evaluation tier.
$N{=}2{,}242$ for every method except Kimi-Audio ($N{=}2{,}206$, marked $\ddagger$;
see Table~\ref{tab:results-category}). Row order within Tier A matches
Table~\ref{tab:results-category} (category-level F1 rank) for easy cross-reference.}
\label{tab:results-finegrained}
\small
\renewcommand{\arraystretch}{1.2}
\begin{tabularx}{\textwidth}{L L C C C C}
  \toprule
  \textbf{Tier} & \textbf{Method} & \textbf{Precision} & \textbf{Recall (TPR)} & \textbf{F1} & \textbf{FNR} \\
  \midrule
  \rowcolor{rowgray}
  & gemini-3.1-pro-preview & 60.7\% & 60.8\% & 56.7\% & 39.2\% \\
  \rowcolor{rowgray} & gemini-2.5-pro         & 51.2\% & 53.3\% & 50.1\% & 46.7\% \\
  \rowcolor{rowgray} & gemini-3.5-flash       & 57.4\% & 52.2\% & 47.9\% & 47.8\% \\
  \rowcolor{rowgray} & kimi-audio-7b-instruct$^\ddagger$ & 38.7\% & 39.6\% & 32.9\% & 60.4\% \\
  \rowcolor{rowgray} \multirow{-5}{*}{A} & gemini-2.5-flash & 38.9\% & 41.2\% & 37.0\% & 58.8\% \\
  \midrule
  \multirow{4}{*}{B} & sslam       & 40.3\% & 39.6\% & 38.0\% & 60.4\% \\
                     & panns       & 37.7\% & 37.0\% & 35.9\% & 63.0\% \\
                     & whisper-at  & 33.6\% & 31.4\% & 30.7\% & 68.6\% \\
                     & yamnet      & 27.8\% & 26.1\% & 23.7\% & 73.9\% \\
  \midrule
  \rowcolor{rowgray}
  C & clap & 61.7\% & 54.2\% & 50.6\% & 45.8\% \\
  \midrule
  D & bat & n/a$^\dagger$ & 35.4\% & n/a$^\dagger$ & 64.6\% \\
  \bottomrule
\end{tabularx}
\end{table}

Two patterns stand out immediately. First, within Tier A, \modelname{gemini-3.1-pro-preview}
leads at both granularities by a comfortable margin (8+ F1 points at the category
level over the next-best Gemini model); \modelname{kimi-audio-7b-instruct}, the one
open-weight model in Tier A, lands fourth of five at both the category level — ahead of
\modelname{gemini-2.5-flash} despite being a much smaller,
openly-released model, but last of five at the fine-grained level, behind all four Gemini models.
Second, the category-to-fine-grained F1 drop is universal and large: every method
loses 25--50 F1 points moving from 11-way to 23-way discrimination, and the ranking
of methods shifts substantially between the two tables (e.g.\ CLAP's fine-grained F1
of 50.6\% is only 6.1 points behind the best Gemini model, despite CLAP never
performing any generative reasoning).

\subsection{Where the Category-to-Fine-Grained Gap Comes From}
\label{sec:gap}

The gap is not diffuse noise; it concentrates in a small number of structurally hard
distinctions, visible consistently across methods and confirmed in the per-class
confusion matrices in Appendix~\ref{app:full-results}:

\begin{itemize}
  \item \textbf{\texttt{Water Running - Distant Source} is essentially unsolvable}
        for all five Tier A models (F1~$\approx$~0 throughout, including
        Kimi-Audio's F1~$=$~0.02), with 57--83\% of these samples predicted as
        \texttt{Water Running - Close Source} across all five (Kimi-Audio: 65\%).
        Models correctly detect running water; they cannot judge recording distance
        from the audio alone.
  \item \textbf{\texttt{Doorbell - Wireless} is similarly unsolvable} for four of
        five Tier A models (all except \modelname{gemini-2.5-flash}), with errors split
        between \texttt{Doorbell - Mechanical} and \texttt{Doorbell - Digital}, and
        11--35\% spilling into \texttt{Appliance Alarm} — a wireless doorbell chime
        is acoustically closer to an electronic alarm tone than to a mechanical
        door-knocker. Kimi-Audio shows the strongest version of this particular
        confusion (35\% of its \texttt{Doorbell - Wireless} predictions land on
        \texttt{Appliance Alarm}, versus 11--19\% for the four Gemini models).
  \item \textbf{Siren sub-types show a model-family default bias.} All five Tier A
        models struggle to separate \texttt{Ambulance}/\texttt{Fire}/\texttt{Police}/
        \texttt{Other}, but \modelname{gemini-3.1-pro-preview} and
        \modelname{gemini-2.5-pro} default toward \textbf{Police} under uncertainty,
        while \modelname{gemini-3.5-flash} and \modelname{gemini-2.5-flash} default
        toward \textbf{Ambulance} — the same underlying acoustic confusion,
        resolved by a different learned prior per model family. Kimi-Audio joins the
        Ambulance-defaulting group but far more extremely: 98\% of its
        \texttt{Siren - Police} samples and 86\% of its \texttt{Siren - Fire} samples
        are predicted as \texttt{Siren - Ambulance}, effectively collapsing all
        siren sub-types into a single answer rather than showing a softer bias.
\end{itemize}

\begin{figure}[H]
  \centering
  \includegraphics[width=0.8\textwidth]{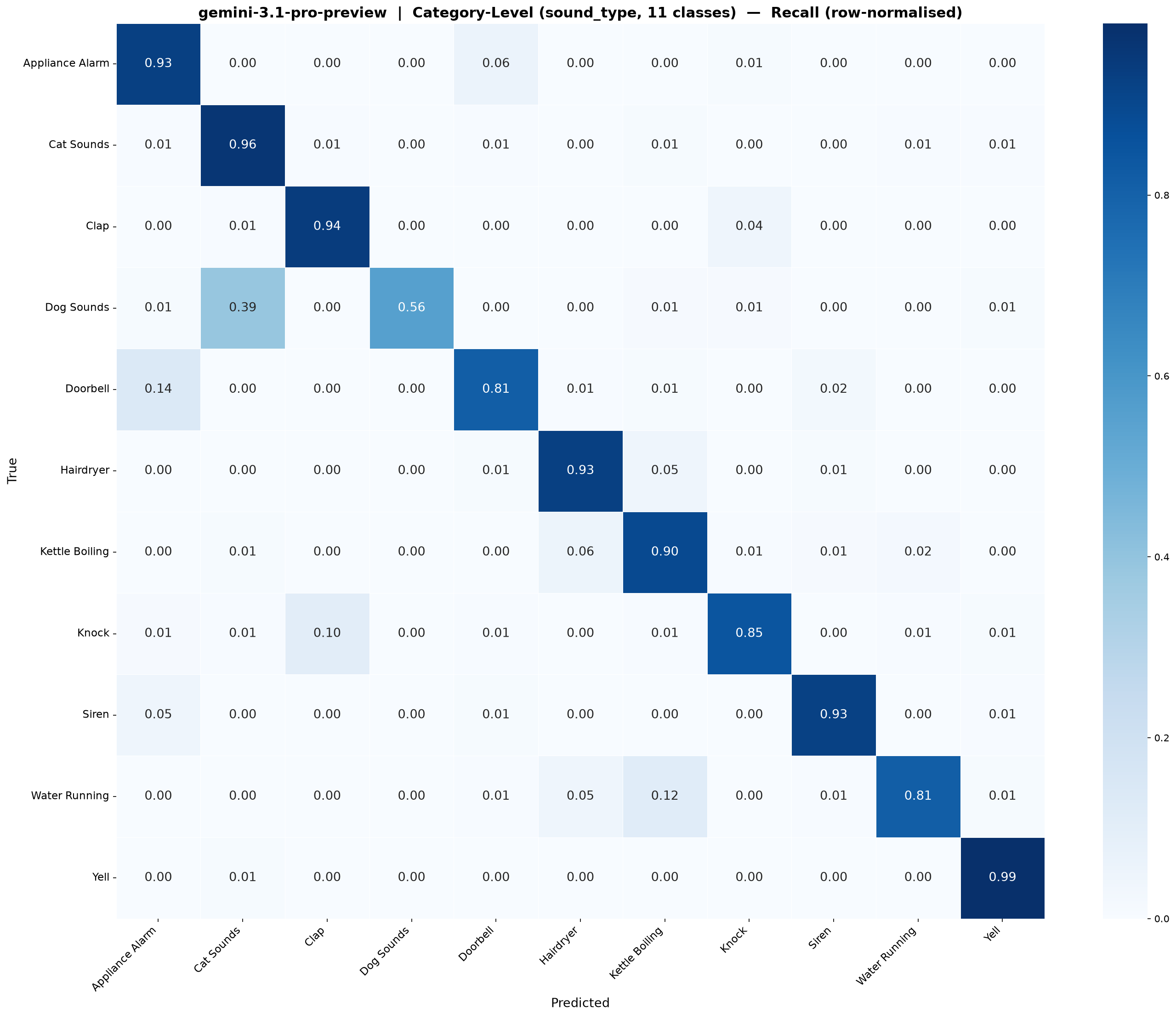}
  \caption{Category-level confusion matrix (row-normalized recall) for
  \modelname{gemini-3.1-pro-preview}, the best-performing Tier A model. The full
  23-class fine-grained matrix, where the confusions above are visible in detail,
  is given in Appendix~\ref{app:full-results} (Figure~\ref{fig:cm-best-fine}).}
  \label{fig:cm-best-cat}
\end{figure}

\subsection{Top-K Accuracy (Tier B)}
\label{sec:topk}

Tables~\ref{tab:results-category} and~\ref{tab:results-finegrained} score every
method on its single best answer (Top-1). Tier B methods are unique among the four
tiers in that they naturally produce a full \emph{ranked} list of candidates rather
than a single forced choice (Tier A and Tier C commit to one answer; Tier D produces
free text with no ranking at all), so we additionally ask: does the true label
appear \emph{anywhere} in the model's top-$K$ ranked, taxonomy-mapped predictions,
for $K \in \{1, 3, 5\}$? $K{=}1$ reproduces the raw accuracy already reported;
$K{=}5$ is the deepest ranking collected at data-preparation time, so it is
the most lenient, best-case figure available from this data.

\begin{table}[H]
\centering
\caption{Top-1/3/5 accuracy for all four Tier B methods, both granularities.
$N{=}2{,}242$ throughout.}
\label{tab:topk-accuracy}
\small
\renewcommand{\arraystretch}{1.2}
\begin{tabularx}{\textwidth}{L C C C}
  \toprule
  \textbf{Method} & \textbf{K=1} & \textbf{K=3} & \textbf{K=5} \\
  \midrule
  \multicolumn{4}{l}{\emph{Category-Level (11 classes)}} \\
  \rowcolor{rowgray}
  sslam       & 88.4\% & 95.9\% & 96.9\% \\
  panns       & 84.3\% & 94.0\% & 96.5\% \\
  \rowcolor{rowgray}
  whisper-at  & 77.8\% & 90.3\% & 93.7\% \\
  yamnet      & 52.1\% & 64.4\% & 66.4\% \\
  \midrule
  \multicolumn{4}{l}{\emph{Fine-Grained (23 classes)}} \\
  \rowcolor{rowgray}
  sslam       & 40.1\% & 50.3\% & 54.2\% \\
  panns       & 37.3\% & 48.8\% & 53.5\% \\
  \rowcolor{rowgray}
  whisper-at  & 31.7\% & 44.5\% & 50.8\% \\
  yamnet      & 26.5\% & 35.1\% & 40.7\% \\
  \bottomrule
\end{tabularx}
\end{table}

The gap between $K{=}1$ and $K{=}5$ is large and consistent: every Tier B method
gains roughly 14--19 Fine-Grained points by $K{=}5$ (e.g.\ SSLAM: 40.1\%
$\rightarrow$ 54.2\%; Whisper-AT: 31.7\% $\rightarrow$ 50.8\%), indicating the correct answer is frequently present in these
models' ranked output even when their single top choice misses. Most of that gain
arrives by $K{=}3$; going from $K{=}3$ to $K{=}5$ adds a further 4--6 points, a
diminishing but still real return. At $K{=}5$, SSLAM's Fine-Grained accuracy
(54.2\%) approaches CLAP's Top-1 figure (54.5\%, Table~\ref{tab:accuracy-full}),
is on par with the strongest non-preview Tier A models (\modelname{gemini-2.5-pro, 54.1\%}) and trails only \modelname{gemini-3.1-pro-preview} (61.5\%) - though this is not an apples-to-apples comparision: $K{=}5$ credits a
method for having the right answer anywhere in five guesses, a substantially easier
bar than the single forced answer every other tier is held to.

For the detailed per-class view, we focus the $K{=}5$ (best-case) confusion
matrices in Appendix~\ref{app:full-results}, alongside the standard Top-1 matrices,
since $K{=}5$ is the most informative depth for characterizing what each Tier B
method's ranked output is actually capable of resolving.

\subsection{Reasoning-Strategy Analysis (Gemini Models)}
\label{sec:reasoning}

\subsubsection{Verbosity does not predict accuracy}

\begin{figure}[H]
  \centering
  \includegraphics[width=0.7\textwidth]{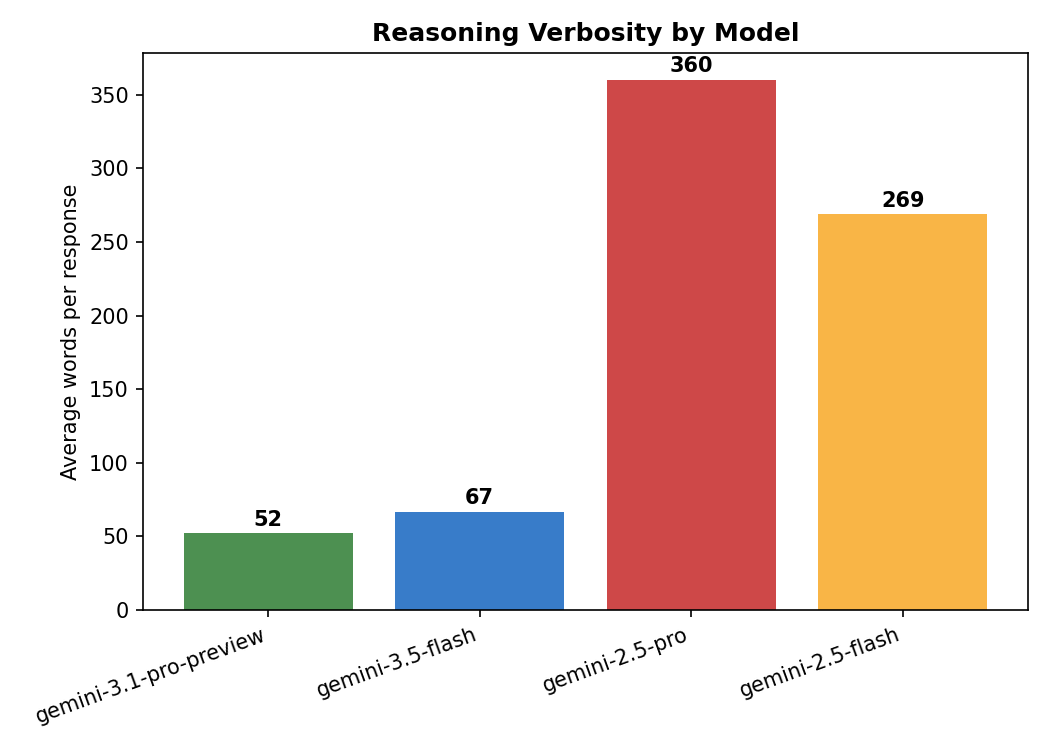}
  \caption{Average chain-of-thought response length by model (same data as
  Table~\ref{tab:verbosity}).}
  \label{fig:word-count}
\end{figure}

Table~\ref{tab:verbosity} and Figure~\ref{fig:word-count} show average chain-of-thought
response length by model. The most accurate model, \modelname{gemini-3.1-pro-preview},
is also the most concise — nearly 7$\times$ shorter than \modelname{gemini-2.5-pro},
which writes exhaustive, structured, multi-step analyses (87.6\% of its responses
use explicit numbered steps, vs.\ 0\% for \modelname{gemini-3.1-pro-preview}) without
converting that extra length into better fine-grained F1 than the far terser
\modelname{gemini-3.5-flash} (Table~\ref{tab:results-finegrained}). Response length and
accuracy are, if anything, inversely related across these four models.

\begin{table}[H]
\centering
\caption{Average chain-of-thought response length by model.}
\label{tab:verbosity}
\small
\begin{tabularx}{0.6\textwidth}{L C}
  \toprule
  \textbf{Model} & \textbf{Avg. words / response} \\
  \midrule
  \rowcolor{rowgray}
  gemini-3.1-pro-preview & 52 \\
  gemini-3.5-flash       & 67 \\
  \rowcolor{rowgray}
  gemini-2.5-flash       & 269 \\
  gemini-2.5-pro         & 360 \\
  \bottomrule
\end{tabularx}
\end{table}

\subsubsection{An apparent strategy-accuracy pattern is a difficulty confound}

\begin{figure}[H]
  \centering
  \includegraphics[width=0.75\textwidth]{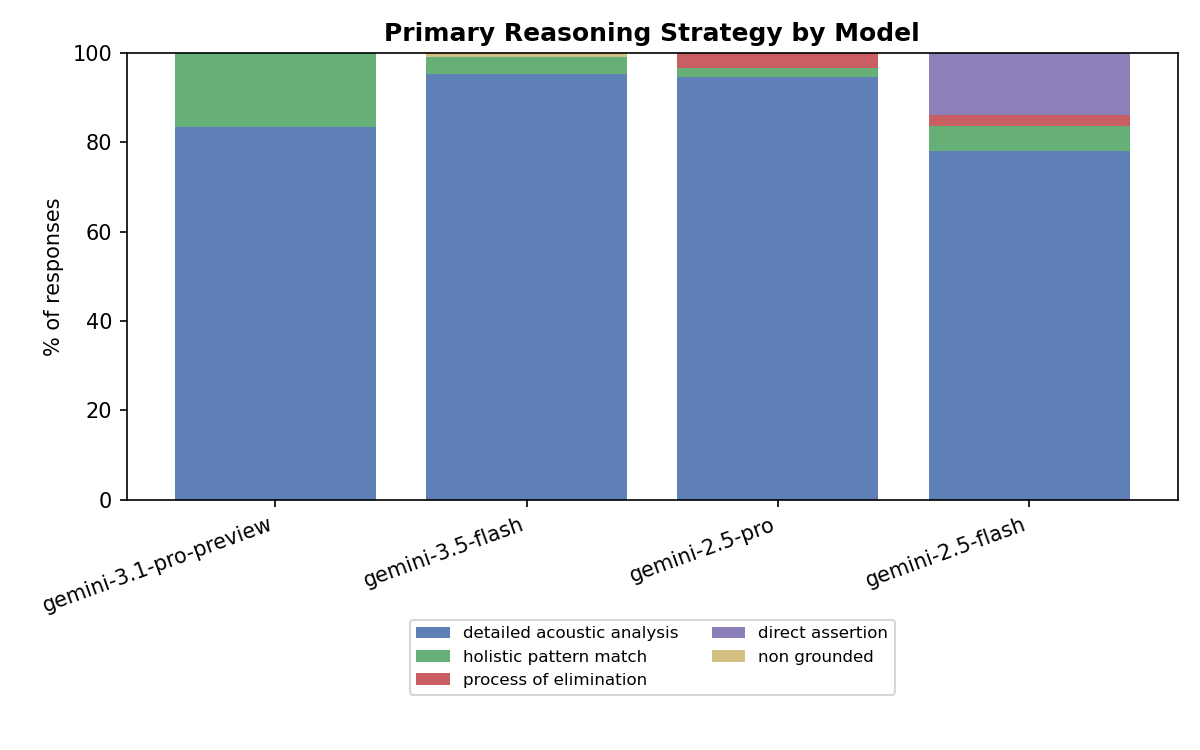}
  \caption{Primary reasoning strategy distribution by model. \emph{Detailed acoustic
  analysis} dominates throughout (78.1--95.2\%); \emph{holistic pattern match} is a
  distant second everywhere, but far more common for
  \modelname{gemini-3.1-pro-preview} (16.5\%) than the other three models
  (1.9--5.5\%).}
  \label{fig:strategy-dist}
\end{figure}

Each response was tagged with a primary reasoning strategy. Figure~\ref{fig:strategy-dist}
shows the distribution: \emph{detailed acoustic analysis} (explicit discussion of
frequency, timbre, envelope, pitch) dominates for every model, from 78.1\%
(\modelname{gemini-2.5-flash}) up to 95.2\% (\modelname{gemini-3.5-flash}) of responses.
\emph{Holistic pattern match} (a quick "this sounds like X" judgment with no feature
breakdown) is the clear second-most-common strategy for all four models, but its
share varies considerably: \modelname{gemini-3.1-pro-preview} uses it in 16.5\% of
responses, three to eight times more often than the other three models
(1.9--5.5\%). The remaining strategies (\emph{process of elimination},
\emph{direct assertion}, \emph{non-grounded}) are minor and inconsistent across
models, appearing in at most 13.9\% of any one model's responses.

\begin{figure}[H]
  \centering
  \includegraphics[width=0.75\textwidth]{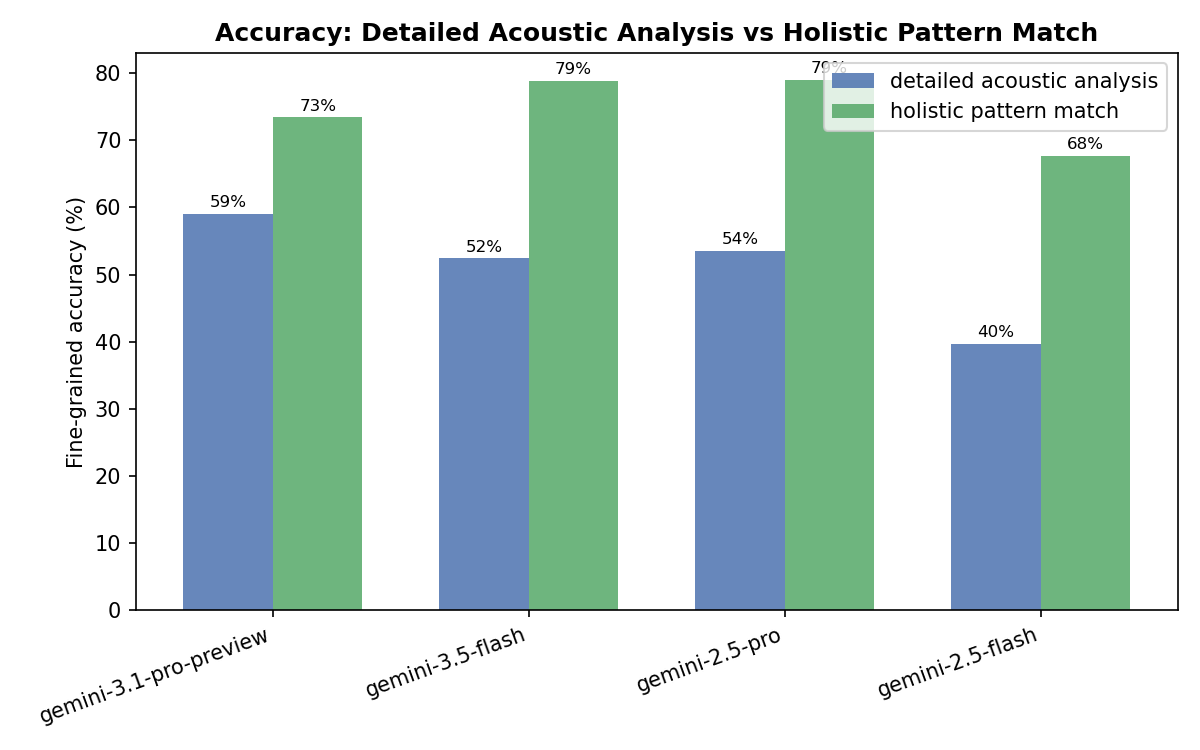}
  \caption{Fine-grained accuracy conditioned on primary reasoning strategy, per
  model (exact values in Table~\ref{tab:strategy-accuracy}). In every model the
  holistic-pattern-match bar is taller than the detailed-acoustic-analysis bar.}
  \label{fig:strategy-acc}
\end{figure}

Naively cross-referencing strategy against fine-grained accuracy
(Table~\ref{tab:strategy-accuracy}, visualized in Figure~\ref{fig:strategy-acc})
shows the same pattern in all four models: responses tagged \emph{holistic pattern
match} are correct 67.7--79.1\% of the time, versus only 39.7--59.1\% for
\emph{detailed acoustic analysis} — a 14.3--28.0 point gap in every model, with no
exceptions.

\begin{table}[H]
\centering
\caption{Fine-grained accuracy by primary reasoning strategy, for the two dominant
strategies, per model. $n$ is the number of responses tagged with that strategy
(out of 2{,}242 per model); the holistic-pattern-match bar is taller than the
detailed-acoustic-analysis bar in every model.}
\label{tab:strategy-accuracy}
\small
\renewcommand{\arraystretch}{1.2}
\begin{tabularx}{\textwidth}{L C C C}
  \toprule
  \textbf{Model} & \textbf{Detailed-analysis acc. (n)} & \textbf{Holistic-match acc. (n)} & \textbf{Gap (points)} \\
  \midrule
  \rowcolor{rowgray}
  gemini-3.1-pro-preview & 59.1\% (n=1{,}871) & 73.4\% (n=369) & 14.3 \\
  gemini-2.5-pro         & 53.5\% (n=2{,}122) & 79.1\% (n=43)  & 25.6 \\
  \rowcolor{rowgray}
  gemini-3.5-flash       & 52.4\% (n=2{,}135) & 78.8\% (n=85)  & 26.4 \\
  gemini-2.5-flash       & 39.7\% (n=1{,}750) & 67.7\% (n=124) & 28.0 \\
  \bottomrule
\end{tabularx}
\end{table}

Read at face value, this looks like "less analysis produces better answers." We do
not believe that is the correct interpretation. Restricting the comparison to
\modelname{gemini-3.1-pro-preview} — the model with by far the highest holistic-match
rate (Figure~\ref{fig:strategy-dist}) — and conditioning on individual classes
(Table~\ref{tab:strategy-class}) shows that the two strategies score about the same
\emph{within} a given class (e.g., 100.0\% vs.\ 98.9\% on \texttt{Yell}); the
apparent strategy effect disappears once class difficulty is held fixed.

\begin{table}[H]
\centering
\caption{Fine-grained accuracy by reasoning strategy, within class, for
\modelname{gemini-3.1-pro-preview}. These three classes account for 47\% of all
\emph{holistic pattern match} responses from this model.}
\label{tab:strategy-class}
\small
\begin{tabularx}{\textwidth}{L C C C}
  \toprule
  \textbf{Class} & \textbf{Overall accuracy} & \textbf{Detailed-analysis acc. (n)} & \textbf{Holistic-match acc. (n)} \\
  \midrule
  \rowcolor{rowgray}
  Yell - Yell                   & 99.0\% & 100.0\% (n=9)  & 98.9\% (n=89) \\
  Appliance Alarm                & 93.0\% & 90.7\% (n=54)  & 95.7\% (n=46) \\
  \rowcolor{rowgray}
  Water Running - Close Source   & 80.0\% & 78.7\% (n=61)  & 82.1\% (n=39) \\
  \bottomrule
\end{tabularx}
\end{table}

The more plausible explanation, consistent with Table~\ref{tab:strategy-class}:
models default to a quick holistic judgment precisely when a sound is unambiguous,
and reach for detailed acoustic analysis when it is genuinely hard to place. The
reasoning strategy is a \emph{symptom} of per-sample difficulty, not a cause of the
outcome — but it remains a useful signal: a model defaulting to detailed analysis is
implicitly flagging "this one was hard."

\subsubsection{Confidence calibration is essentially absent}

\begin{figure}[H]
  \centering
  \includegraphics[width=0.7\textwidth]{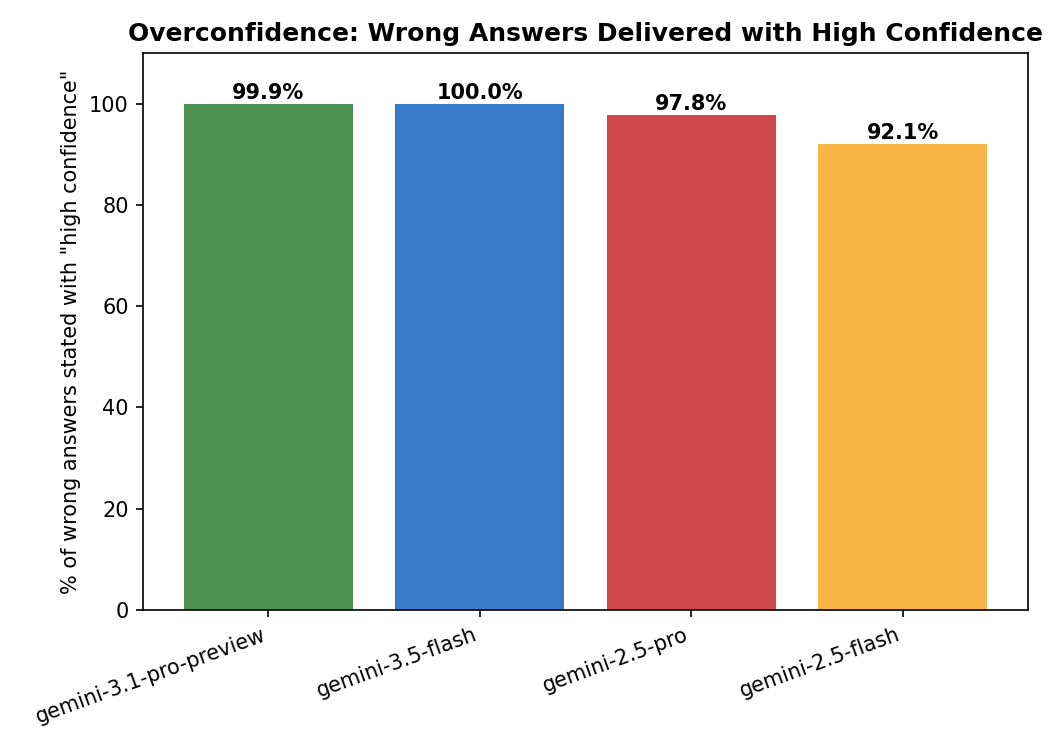}
  \caption{Share of wrong answers still stated with high-confidence language, by
  model (same data as Table~\ref{tab:overconfidence}).}
  \label{fig:overconfidence}
\end{figure}

We measured how often a model's language reads as confident even when its final
answer is wrong (Table~\ref{tab:overconfidence}, Figure~\ref{fig:overconfidence}).
Confidence level is inferred purely from phrasing (e.g., "clearly," "definitely"
vs.\ "possibly," "might be") by the second-pass classifier — models were never asked
to self-report a confidence score.

\begin{table}[H]
\centering
\caption{Share of \emph{wrong} fine-grained answers still stated with high-confidence
language.}
\label{tab:overconfidence}
\small
\begin{tabularx}{0.7\textwidth}{L C}
  \toprule
  \textbf{Model} & \textbf{Wrong answers stated with high-confidence language} \\
  \midrule
  \rowcolor{rowgray}
  gemini-3.5-flash       & 100.0\% \\
  gemini-3.1-pro-preview & 99.9\% \\
  \rowcolor{rowgray}
  gemini-2.5-pro         & 97.8\% \\
  gemini-2.5-flash       & 92.1\% \\
  \bottomrule
\end{tabularx}
\end{table}

As Table~\ref{tab:overconfidence} and Figure~\ref{fig:overconfidence} show, all four
models are essentially uniform on this measure — the range spans only 92.1\% to
100.0\%, with no model showing meaningfully better calibration than the others.
Hedging language ("possibly," "might," "uncertain") appears less than once per
response on average for every model. There is effectively no verbal signal in these
chain-of-thought traces that reliably distinguishes a correct answer from an
incorrect one; a system that needs to know when to defer to a human cannot obtain
that signal from a model's stated confidence on this task.

\section{Discussion}
\label{sec:discussion}

\subsection{What Can and Cannot Be Compared Directly}

The four-tier structure in Section~\ref{sec:methodology} exists because a single
flat ranking across all eleven methods would overstate precision. The defensible
comparisons are \emph{within} a tier: which Tier A method is best (four Gemini
models plus Kimi-Audio, all given the identical closed-set prompt), and which
fixed-vocabulary tagger is best (Tier B). Tier C (CLAP) and Tier D (BAT) are each a
category of one in this study and are reported for directional context rather
than as ranked competitors.

That said, one cross-tier observation is robust enough to be worth surfacing
directionally: at the category level, task-agnostic methods that never saw the
candidate list (SSLAM, PANNs in Tier B; CLAP in Tier C) match or exceed the best
task-aware Tier A model's F1, and PANNs approches it. At the fine-grained level, this reverses — the best
Tier A model and CLAP (Tier C) pull ahead of every Tier B tagger by 12+ F1 points.
A plausible reading is that coarse categorization is largely an acoustic
pattern-matching problem, at which purpose-built audio encoders (with or without a
candidate list) are already strong, while fine-grained sub-type disambiguation
increasingly draws on the kind of contextual, elimination-based reasoning that Tier
A's language-generation step provides and Tier B's fixed classification head does
not. We stress this is a hypothesis suggested by the data, not a controlled causal
claim, since Tier A and Tier B differ in task framing as well as in architecture.

Even within Tier A, the five methods are not architecturally uniform: the four
Gemini models are large proprietary systems of undisclosed scale, while
Kimi-Audio-7B-Instruct is a 7B-parameter open-weight model. That it lands
mid-pack in Tier A rather than at the bottom is itself informative — task-aware
closed-set prompting appears to narrow, though not eliminate, the gap between an
open 7B audio-native model and much larger proprietary multimodal systems on this
task — but its lower reasoning-instruction compliance (Section~\ref{sec:methodology})
means its Tier A ranking reflects both classification ability and instruction-
following, in a way the four Gemini models' more consistent compliance does not.

\subsection{Practical Guidance}

For a use case that only needs the broad category (is this a siren or a doorbell?),
several method families are viable, and a task-agnostic tagger (SSLAM, PANNs) or a
zero-shot model (CLAP) may be sufficient without needing an LLM call per sample. For
a use case needing the exact sub-type, the strongest current methods top out around
51--57\% F1, and Tier A (task-aware LLM with an explicit candidate list) or Tier C (CLAP) are the
strongest available options; expect specific, predictable failure modes — recording-distance
and directionality judgments, and acoustically adjacent sub-types within a
category — rather than random error. If an open-weight, self-hostable option is a
requirement, Kimi-Audio-7B-Instruct is a usable Tier A alternative to the Gemini
models, at a cost in fine-grained F1 (32.9\% vs.\ 56.7\% for the best Gemini model)
and in reliability (1.6\% of samples yield no scoreable answer at all, versus 0\%
for the Gemini models), both worth budgeting for in a production deployment.

\subsection{Limitations}

\begin{itemize}
  \item \textbf{Tier C's task framing partially overlaps Tier A's.} CLAP is given
        the same 23 candidates as the Gemini models (as text), which is part of why
        its numbers are more directly comparable to Tier A than Tier B's are; this
        is noted explicitly rather than treated as a hidden asymmetry, but it means
        CLAP's strong fine-grained score should not be read as evidence that
        zero-shot similarity generally rivals LLM reasoning on unconstrained audio
        tagging.
  \item \textbf{Tier D's scoring depends on a judge model.} BAT's free-text answers
        are graded by \modelname{gemini-2.5-flash-lite} rather than exact match; judge
        leniency or strictness is a source of variance this study does not
        independently calibrate against a second judge or human raters.
  \item \textbf{Tier B's mapping is post-hoc.} The AudioSet-to-taxonomy mapping used
        for YAMNet, PANNs, Whisper-AT, and SSLAM is hand-constructed and tagged
        exact/approximate at data-preparation time; a different, equally reasonable
        mapping could shift these numbers somewhat, though the relative ordering
        within Tier B was stable under manual review.
  \item \textbf{BAT's spatial-audio design is not exercised here.} BAT's Spatial-AST
        encoder expects binaural input with directional cues; our clips are
        effectively mono, so the channel was duplicated rather than fabricating
        synthetic spatial cues. This is an honest "no directional information"
        representation but means BAT is evaluated outside its primary designed use
        case.
  \item \textbf{Kimi-Audio's 36 excluded samples are a genuine reliability
        difference, not a scoring artifact.} We confirmed this by re-attempting
        extraction with three successive rounds of parser improvements and by
        re-running the 36 samples under identical settings, which reproduced the
        same failures rather than resolving them (Kimi-Audio's text decoding is
        greedy, so a same-settings retry has limited ability to escape a
        repetition-loop or refusal outcome). We did not pursue a targeted fix
        (e.g., raising the text repetition penalty, currently uninformatively set
        to 1.0) specifically for these samples, since doing so would apply
        different generation settings to a subset of Kimi-Audio's data than to the
        rest, undermining the like-for-like comparison this study is built around.
  \item \textbf{Single dataset, single run.} All numbers reflect one balanced
        23-class dataset and a single evaluation pass per method (no repeated
        sampling or multiple-seed variance estimates), and the 11-category taxonomy
        reflects the specific sound sources of interest to this study rather than a
        general-purpose acoustic scene ontology.
\end{itemize}

\subsection{Future Work}
\label{sec:future-work}

Natural extensions include: a second, independent LLM judge (or a small human-rated
subset) to quantify Tier D's judge-variance; testing whether few-shot prompting
can induce BAT toward closed-set letter selection rather than its default binary
yes/no bias; and extending the tiered framework to acoustic scenes with genuine
spatial/directional structure, which would let BAT's native architecture be
evaluated on-task rather than degraded to mono input.

A further direction motivated by concurrent work (Section~\ref{sec:background}) is
to manipulate the audio front end itself as an independent variable within Tier A.
All Tier A methods in this study are encoder-based to some degree — Kimi-Audio
explicitly tokenizes audio through a pretrained Whisper-derived encoder before its
LLM core ever sees it, and Gemini's proprietary audio pathway is almost certainly
similar in spirit. Fan~et~al.~\cite{fan2026llmreadspectrogramencoderfree} report that removing this encoder
entirely (projecting Mel-spectrogram patches directly into the LLM) improves
performance specifically on tasks carried by low-level acoustic structure,
including environmental sound classification, at some cost to knowledge-intensive
tasks. Our own fine-grained results are dominated by exactly this kind of
acoustic discrimination — distinguishing \texttt{Siren - Police} from
\texttt{Siren - Ambulance}, or a wireless doorbell chime from a mechanical one,
plausibly turns on preserved pitch contour and timbre rather than semantic
content. Whether an encoder-free architecture narrows Kimi-Audio's fine-grained
gap to the Gemini models specifically on these acoustically-driven distinctions,
or whether Gemini's advantage instead reflects sheer model scale or training data,
is a testable question our tiered framework is directly set up to answer, given
access to a Mel-LLM-style checkpoint evaluated under the same closed-set prompt.

\section{Conclusion}
\label{sec:conclusion}

Comparing eleven audio classification methods spanning four fundamentally different
evaluation paradigms on the identical 2{,}242-sample, 23-class dataset shows that no
single method dominates once category and fine-grained granularity, and task framing,
are both accounted for. Task-aware closed-set LLM selection
(\modelname{gemini-3.1-pro-preview}) gives the best fine-grained F1 (56.7\%) of any
method tested, but task-agnostic fixed-vocabulary taggers and zero-shot audio-text
similarity are competitive or superior at the coarser category level despite never
seeing the candidate list. The open-weight Kimi-Audio-7B-Instruct, evaluated under
the identical closed-set prompt, reaches a competitive 67.5\%/32.9\% category/fine-
grained F1 for its size, though with a measurably higher rate of unscoreable
responses (1.6\% of samples) than the four Gemini models (0\%), reflecting less
consistent instruction-following rather than a scoring artifact. The
category-to-fine-grained accuracy gap is concentrated in a small number of
structurally hard, consistent failure modes rather than diffuse noise, and these
generalize to Kimi-Audio, in some cases (Siren sub-types) more extremely than any
Gemini model. Separately, analyzing the Gemini models' own chain-of-thought reveals
that more reasoning text does not buy more accuracy, that an apparent
reasoning-strategy effect on accuracy is confounded by per-sample difficulty, and —
most consequential for downstream use — that these models give almost no usable
signal, in their stated confidence, of when they are wrong.


\section*{Ethics Statement}

This study evaluates model performance on a curated set of short environmental sound
recordings (e.g., doorbells, sirens, appliance sounds) and involves no human
subjects, personal data, or sensitive content.

\bibliographystyle{plain}
\bibliography{references}

\appendix

\section{The 23 Fine-Grained Candidate Classes}
\label{app:candidates}

Table~\ref{tab:candidates} lists the exact 23 lettered candidates given to every
Tier A (Gemini) model for every sample, and to Tier C (CLAP) as natural-language
phrasings of the same 23 classes. These collapse into the 11 broader Category-Level
classes: Appliance Alarm, Cat Sounds, Clap, Dog Sounds, Doorbell, Hairdryer, Kettle
Boiling, Knock, Siren, Water Running, Yell.

\begin{table}[H]
\centering
\caption{The 23 fine-grained candidate classes.}
\label{tab:candidates}
\small
\begin{tabularx}{0.7\textwidth}{C L L}
  \toprule
  \textbf{Letter} & \textbf{sound\_type} & \textbf{sub\_category} \\
  \midrule
  \rowcolor{rowgray} A & Appliance Alarm & Appliance Alarm \\
  B & Cat Sounds & Meow \\
  \rowcolor{rowgray} C & Cat Sounds & Purr \\
  D & Clap & Double Clap \\
  \rowcolor{rowgray} E & Clap & Single Clap \\
  F & Clap & Triple Clap \\
  \rowcolor{rowgray} G & Dog Sounds & Growl \\
  H & Dog Sounds & Howl \\
  \rowcolor{rowgray} I & Doorbell & Digital \\
  J & Doorbell & Mechanical \\
  \rowcolor{rowgray} K & Doorbell & Wireless \\
  L & Hairdryer & Hair Dryer \\
  \rowcolor{rowgray} M & Kettle Boiling & Boiling \\
  N & Kettle Boiling & Whistling \\
  \rowcolor{rowgray} O & Knock & Door Knock \\
  P & Knock & Surface Knock \\
  \rowcolor{rowgray} Q & Siren & Ambulance \\
  R & Siren & Fire \\
  \rowcolor{rowgray} S & Siren & Other \\
  T & Siren & Police \\
  \rowcolor{rowgray} U & Water Running & Close Source \\
  V & Water Running & Distant Source \\
  \rowcolor{rowgray} W & Yell & Yell \\
  \bottomrule
\end{tabularx}
\end{table}

\section{Exact Prompts Used}
\label{app:prompts}

This appendix gives the verbatim text prompts used for every method that consumes
a text input (Tier A, Tier C, and Tier D; Tier B methods receive no text prompt at
all and are omitted). All prompts are reproduced exactly as sent, with no
paraphrasing.

\subsection{Tier A: Candidate-Selection Prompt (Gemini and Kimi-Audio)}
\label{app:prompt-tiera}

The identical prompt below was used for all four Gemini models and for
Kimi-Audio-7B-Instruct (Section~\ref{sec:methodology}). \texttt{\{candidate\_list\}}
is substituted with the 23-line lettered list given in
Appendix~\ref{app:candidates}.

\begin{quote}
\small\ttfamily
Listen carefully to the audio file provided.\\[4pt]
Tell me what you hear. You are free to use any signal processing capability or
internal sound classification system you may have access to. Tell me what
approach you used, and share your full chain of thought.\\[4pt]
Choose the single candidate below that best identifies the true source of the
sound. You MUST pick exactly one letter from this list (use the exact letter):\\
\{candidate\_list\}\\[4pt]
After your analysis, provide your final answer on the very last line in this
exact JSON format (no markdown, no extra text after it):\\
\{"answer": "X"\}\\[4pt]
Where X is the single capital letter of your chosen candidate.
\end{quote}

\subsection{Tier C: CLAP Candidate Phrasings}
\label{app:prompt-tierc}

CLAP does not receive a generative prompt; instead, each of the 23 candidates
(same letter assignment as Appendix~\ref{app:candidates}) is given to CLAP as a
short natural-language phrase, embedded once, and compared against the audio
embedding via cosine similarity.

\begin{table}[H]
\centering
\caption{The 23 candidate phrasings given to CLAP.}
\label{tab:clap-phrasings}
\small
\begin{tabularx}{\textwidth}{C L L L}
  \toprule
  \textbf{Letter} & \textbf{sound\_type} & \textbf{sub\_category} & \textbf{CLAP phrase} \\
  \midrule
  \rowcolor{rowgray} A & Appliance Alarm & Appliance Alarm & the sound of an appliance alarm beeping \\
  B & Cat Sounds & Meow & a cat meowing \\
  \rowcolor{rowgray} C & Cat Sounds & Purr & a cat purring \\
  D & Clap & Double Clap & two hands clapping twice in quick succession \\
  \rowcolor{rowgray} E & Clap & Single Clap & a single hand clap \\
  F & Clap & Triple Clap & hands clapping three times in quick succession \\
  \rowcolor{rowgray} G & Dog Sounds & Growl & a dog growling \\
  H & Dog Sounds & Howl & a dog howling \\
  \rowcolor{rowgray} I & Doorbell & Digital & a digital doorbell chime \\
  J & Doorbell & Mechanical & a mechanical doorbell ringing \\
  \rowcolor{rowgray} K & Doorbell & Wireless & a wireless doorbell chime \\
  L & Hairdryer & Hair Dryer & a hair dryer blowing \\
  \rowcolor{rowgray} M & Kettle Boiling & Boiling & water boiling in a kettle \\
  N & Kettle Boiling & Whistling & a kettle whistling \\
  \rowcolor{rowgray} O & Knock & Door Knock & someone knocking on a door \\
  P & Knock & Surface Knock & someone knocking on a surface \\
  \rowcolor{rowgray} Q & Siren & Ambulance & an ambulance siren \\
  R & Siren & Fire & a fire truck siren \\
  \rowcolor{rowgray} S & Siren & Other & a generic emergency siren \\
  T & Siren & Police & a police car siren \\
  \rowcolor{rowgray} U & Water Running & Close Source & water running close to the microphone \\
  V & Water Running & Distant Source & water running far from the microphone \\
  \rowcolor{rowgray} W & Yell & Yell & a person yelling \\
  \bottomrule
\end{tabularx}
\end{table}

\subsection{Tier D: BAT's Open-Vocabulary Question}
\label{app:prompt-tierd}

Unlike Tier A, BAT receives no candidate list and no format constraint. The entire
prompt is a fixed, generic question, identical for every sample:

\begin{quote}
\small\ttfamily
Enumerate the sound occurrences in the audio clip.
\end{quote}

\subsection{Tier D: LLM Judge Prompt}
\label{app:prompt-judge}

BAT's free-text answers are not scored by exact match; they are graded by an
independent LLM judge (\modelname{gemini-2.5-flash-lite}), given the true label and
BAT's raw answer, with the following prompt:

\begin{quote}
\small\ttfamily
You are evaluating whether an audio model's free-text description of a sound
correctly identifies its true source.\\[4pt]
True label:\\
\hspace*{1em}sound\_type = "\{true\_sound\_type\}"\\
\hspace*{1em}sub\_category = "\{true\_sub\_category\}"\\[4pt]
The model's free-text answer:\\
---\\
\{answer\}\\
---\\[4pt]
Does this answer correctly identify the sound as being a "\{true\_sound\_type\}",
specifically "\{true\_sub\_category\}"? Consider synonyms and paraphrases.
Judge category-level (broad sound\_type match) and fine-grained (specific
sub\_category match) separately.\\[4pt]
Respond with ONLY this JSON on the last line (no markdown):\\
\{"category\_correct": true\_or\_false, "finegrained\_correct": true\_or\_false,
"reasoning": "one sentence"\}
\end{quote}

\section{Complete AudioSet-to-Taxonomy Mapping (Tier B)}
\label{app:taxonomy-mapping}

Table~\ref{tab:taxonomy-mapping} lists every entry in the lookup table used to translate
YAMNet/PANNs/Whisper-AT/SSLAM's native AudioSet labels into our 23-class taxonomy,
as described in Section~\ref{sec:match-quality}. Sub-category ``---'' denotes a
category-only entry (the class collapses to Category-Level only; see
Section~\ref{sec:methodology}).

\begin{table}[H]
\centering
\caption{Complete AudioSet-to-taxonomy mapping, all 22 entries.}
\label{tab:taxonomy-mapping}
\small
\begin{tabular}{P{4cm} P{2.15cm} P{2.5cm} P{1.9cm} P{3.2cm}}
  \toprule
  \textbf{AudioSet label} & \textbf{sound\_type} & \textbf{sub\_category} & \textbf{Quality} & \textbf{Reason (if approximate)} \\
  \midrule
  \rowcolor{rowgray} \texttt{"Alarm"} & Appliance Alarm & Appliance Alarm & approximate & Generic parent class; also covers car/smoke/clock alarms \\
  \texttt{"Meow"} & Cat Sounds & Meow & exact & --- \\
  \rowcolor{rowgray} \texttt{"Purr"} & Cat Sounds & Purr & exact & --- \\
  \texttt{"Clapping"} & Clap & --- & exact & AudioSet has no clap-count distinction (single/double/triple) \\
  \rowcolor{rowgray} \texttt{"Growling"} & Dog Sounds & Growl & exact & --- \\
  \texttt{"Howl"} & Dog Sounds & Howl & exact & --- \\
  \rowcolor{rowgray} \texttt{"Bark"} & Dog Sounds & --- & approximate & Generic bark, not specific to growl/howl \\
  \texttt{"Doorbell"} & Doorbell & --- & exact & AudioSet has no digital, mechanical, or wireless distinction \\
  \rowcolor{rowgray} \texttt{"Hair dryer"} & Hairdryer & Hair Dryer & exact & --- \\
  \texttt{"Boiling"} & Kettle Boiling & Boiling & exact & --- \\
  \rowcolor{rowgray} \texttt{"Steam whistle"} & Kettle Boiling & Whistling & exact & --- \\
  \texttt{"Whistling"} & Kettle Boiling & Whistling & approximate & Generic whistling, not specific to a boiling kettle \\
  \rowcolor{rowgray} \texttt{"Knock"} & Knock & --- & exact & AudioSet has no door/surface distinction \\
  \texttt{"Police car (siren)"} & Siren & Police & exact & --- \\
  \rowcolor{rowgray} \texttt{"Ambulance (siren)"} & Siren & Ambulance & exact & --- \\
  \texttt{"Fire engine, fire truck (siren)"} & Siren & Fire & exact & --- \\
  \rowcolor{rowgray} \texttt{"Siren"} & Siren & Other & approximate & Generic siren, not vehicle-type-specific \\
  \texttt{"Civil defense siren"} & Siren & Other & approximate & A distinct siren type with no closer match available \\
  \rowcolor{rowgray} \texttt{"Water tap, faucet"} & Water Running & --- & exact & AudioSet has no recording-distance distinction \\
  \texttt{"Yell"} & Yell & Yell & exact & --- \\
  \rowcolor{rowgray} \texttt{"Shout"} & Yell & Yell & approximate & Near-synonym, not an identical acoustic concept \\
  \texttt{"Bellow"} & Yell & Yell & approximate & Near-synonym, not an identical acoustic concept \\
  \bottomrule
\end{tabular}
\end{table}

\section{Complete Results: Accuracy, Per-Class Metrics, and Confusion Matrices}
\label{app:full-results}

\subsection{Raw Accuracy (for reference)}

Table~\ref{tab:accuracy-full} reports raw accuracy for all eleven methods, for direct
comparability with the project's earlier accuracy-based reporting. As discussed in
Section~\ref{sec:methodology}, we treat macro F1/TPR/FNR (Tables~\ref{tab:results-category}
and~\ref{tab:results-finegrained}) as the primary metrics; accuracy is provided here
as a secondary, widely-recognized reference point.

\begin{table}[H]
\centering
\caption{Raw accuracy, all eleven methods, both granularities. $N{=}2{,}242$ for
every method except Kimi-Audio ($N{=}2{,}206$, marked $\ddagger$).}
\label{tab:accuracy-full}
\small
\begin{tabularx}{\textwidth}{L L C C}
  \toprule
  \textbf{Tier} & \textbf{Method} & \textbf{Category Acc.} & \textbf{Fine-Grained Acc.} \\
  \midrule
  \rowcolor{rowgray}
  A & gemini-3.1-pro-preview & 86.8\% & 61.5\% \\
  A & gemini-2.5-pro         & 78.5\% & 54.1\% \\
  \rowcolor{rowgray}
  A & gemini-3.5-flash       & 78.4\% & 53.4\% \\
  A & kimi-audio-7b-instruct$^\ddagger$ & 71.8\% & 40.8\% \\
  \rowcolor{rowgray}
  A & gemini-2.5-flash       & 68.6\% & 42.0\% \\
  B & sslam                  & 88.4\% & 40.1\% \\
  \rowcolor{rowgray}
  B & panns                  & 84.3\% & 37.3\% \\
  B & whisper-at             & 77.8\% & 31.7\% \\
  \rowcolor{rowgray}
  B & yamnet                 & 52.1\% & 26.5\% \\
  C & clap                   & 89.1\% & 54.5\% \\
  \rowcolor{rowgray}
  D & bat (judge, micro)     & 62.4\% & 35.9\% \\
  \bottomrule
\end{tabularx}
\end{table}

\subsection{Confusion Matrices}

Figures~\ref{fig:cm-a1}--\ref{fig:cm-a20} give the row-normalized (recall) confusion
matrix for every method with a discrete predicted class (all methods except BAT,
Tier D, for which no confusion matrix can be constructed — see
Section~\ref{sec:methodology}), at both the category and fine-grained level.

\begin{figure}[H]
  \centering
  \includegraphics[width=0.85\textwidth]{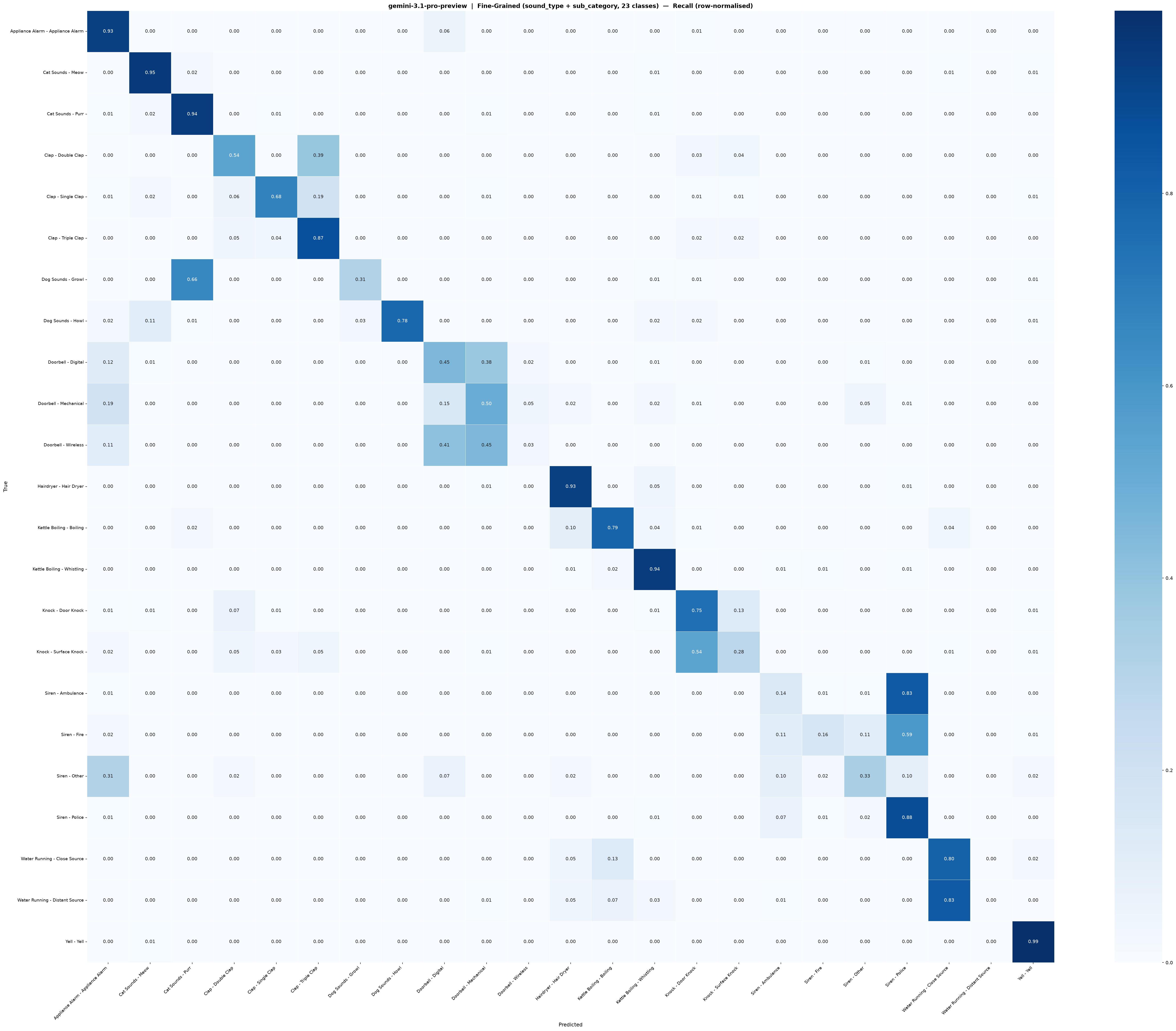}
  \caption{Fine-grained (23-class) confusion matrix (row-normalized recall) for
  \modelname{gemini-3.1-pro-preview}, the best-performing Tier A model. This is the
  detailed view behind the class-level failures discussed in Section~\ref{sec:gap}.}
  \label{fig:cm-best-fine}
\end{figure}

\begin{figure}[H]
  \centering
  \includegraphics[width=0.8\textwidth]{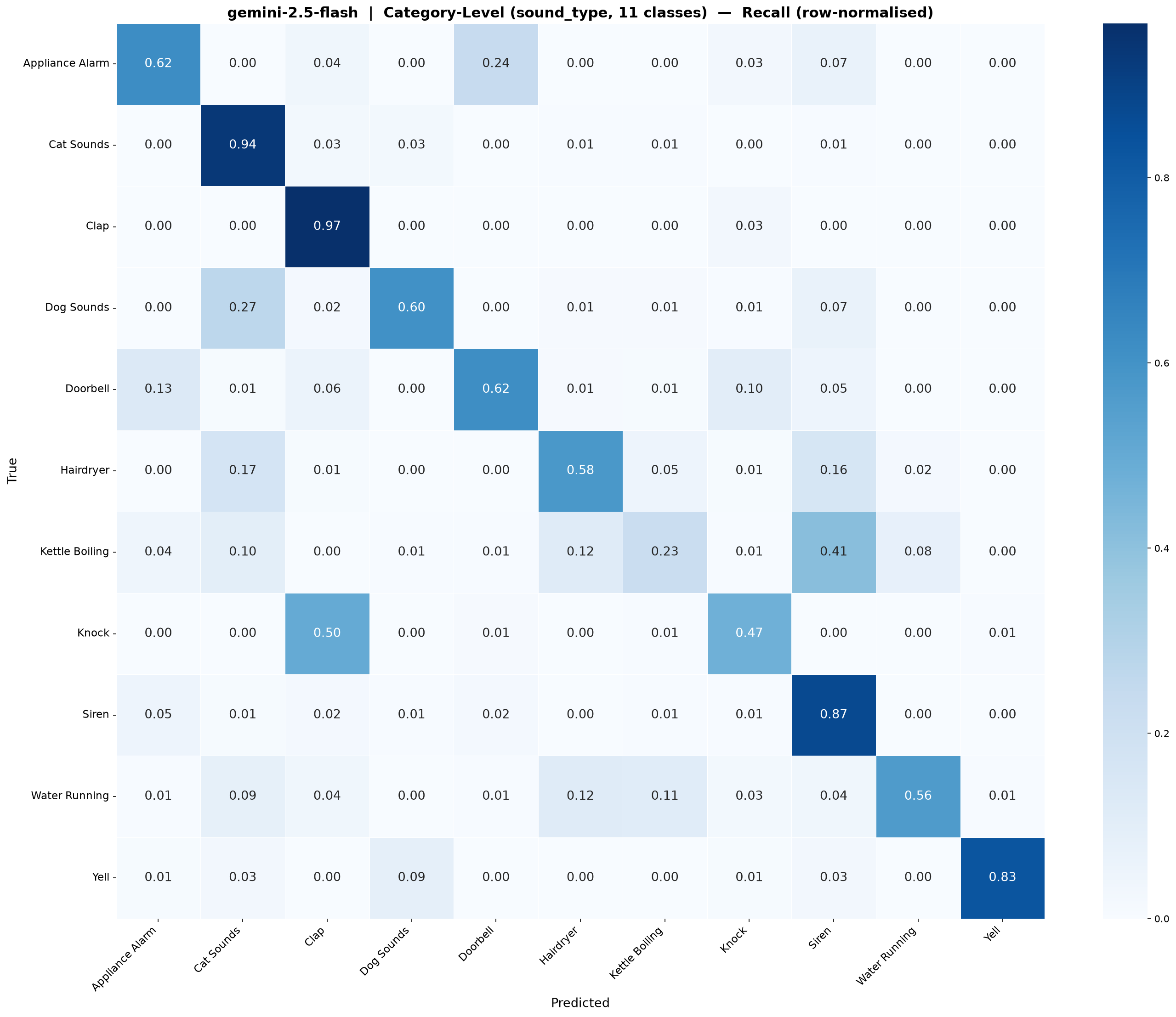}
  \caption{gemini-2.5-flash --- category-level confusion matrix (recall).}
  \label{fig:cm-a1}
\end{figure}
\begin{figure}[H]
  \centering
  \includegraphics[width=0.85\textwidth]{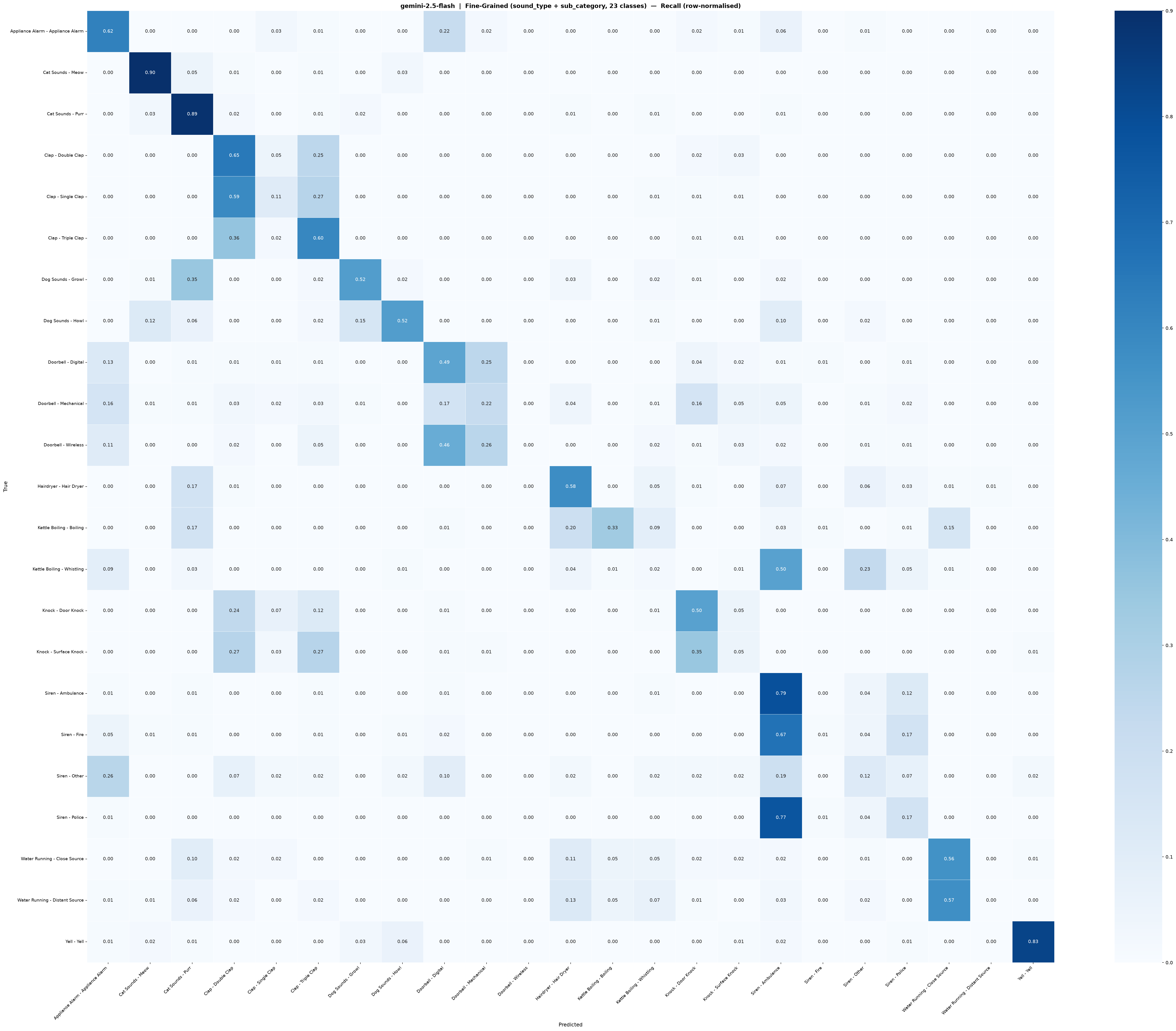}
  \caption{gemini-2.5-flash --- fine-grained confusion matrix (recall).}
\end{figure}

\begin{figure}[H]
  \centering
  \includegraphics[width=0.8\textwidth]{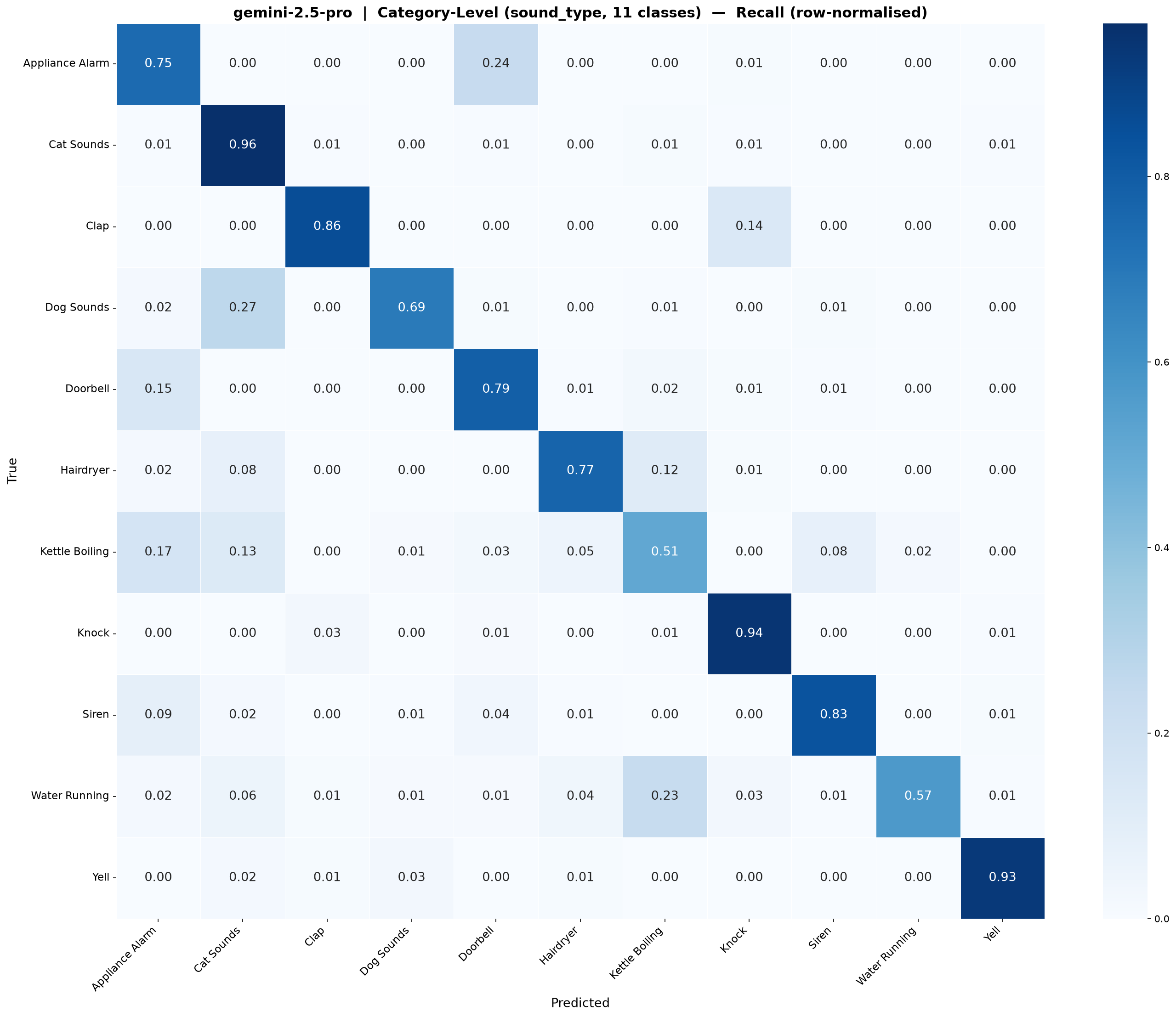}
  \caption{gemini-2.5-pro --- category-level confusion matrix (recall).}
\end{figure}
\begin{figure}[H]
  \centering
  \includegraphics[width=0.85\textwidth]{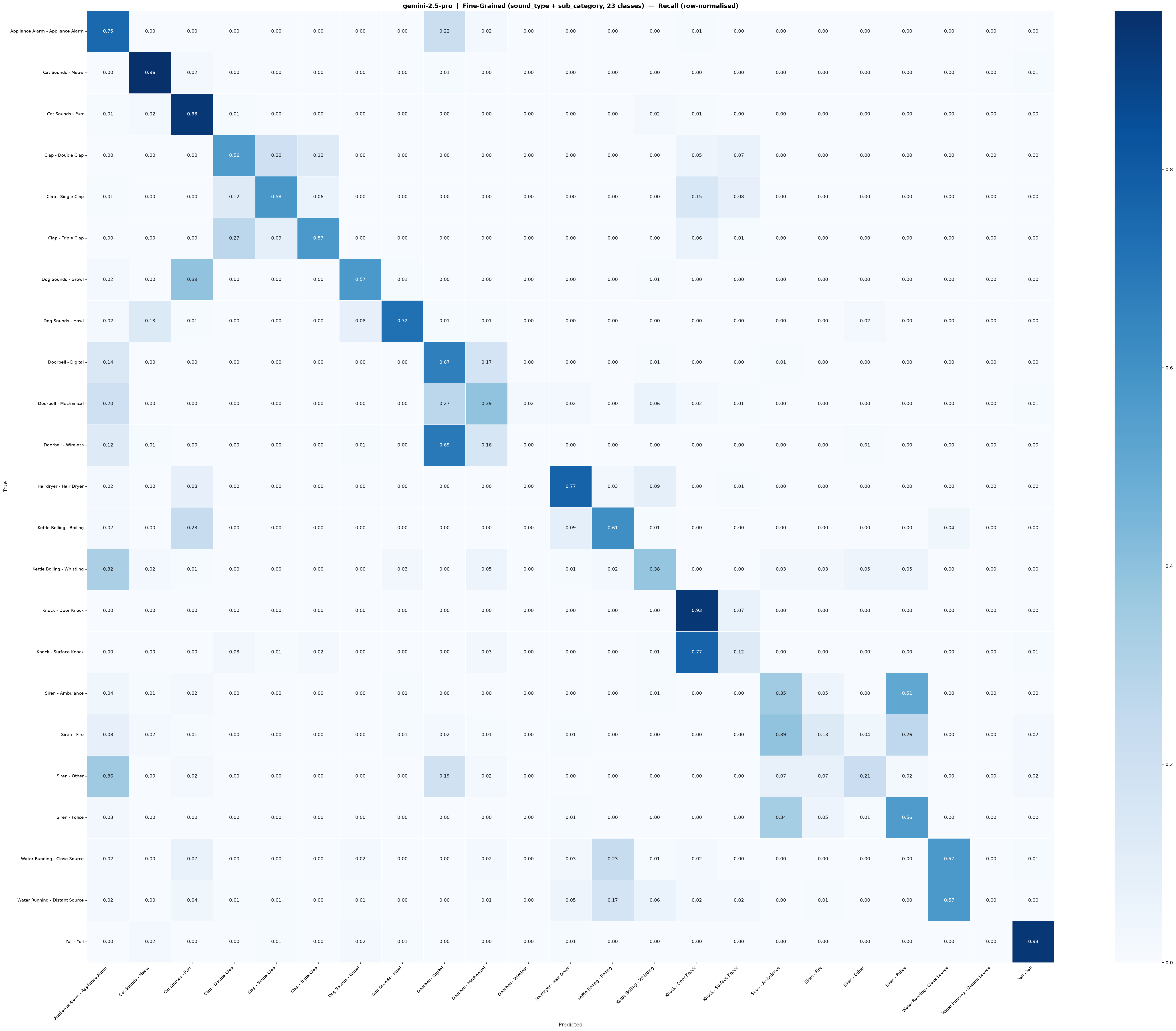}
  \caption{gemini-2.5-pro --- fine-grained confusion matrix (recall).}
\end{figure}

\begin{figure}[H]
  \centering
  \includegraphics[width=0.8\textwidth]{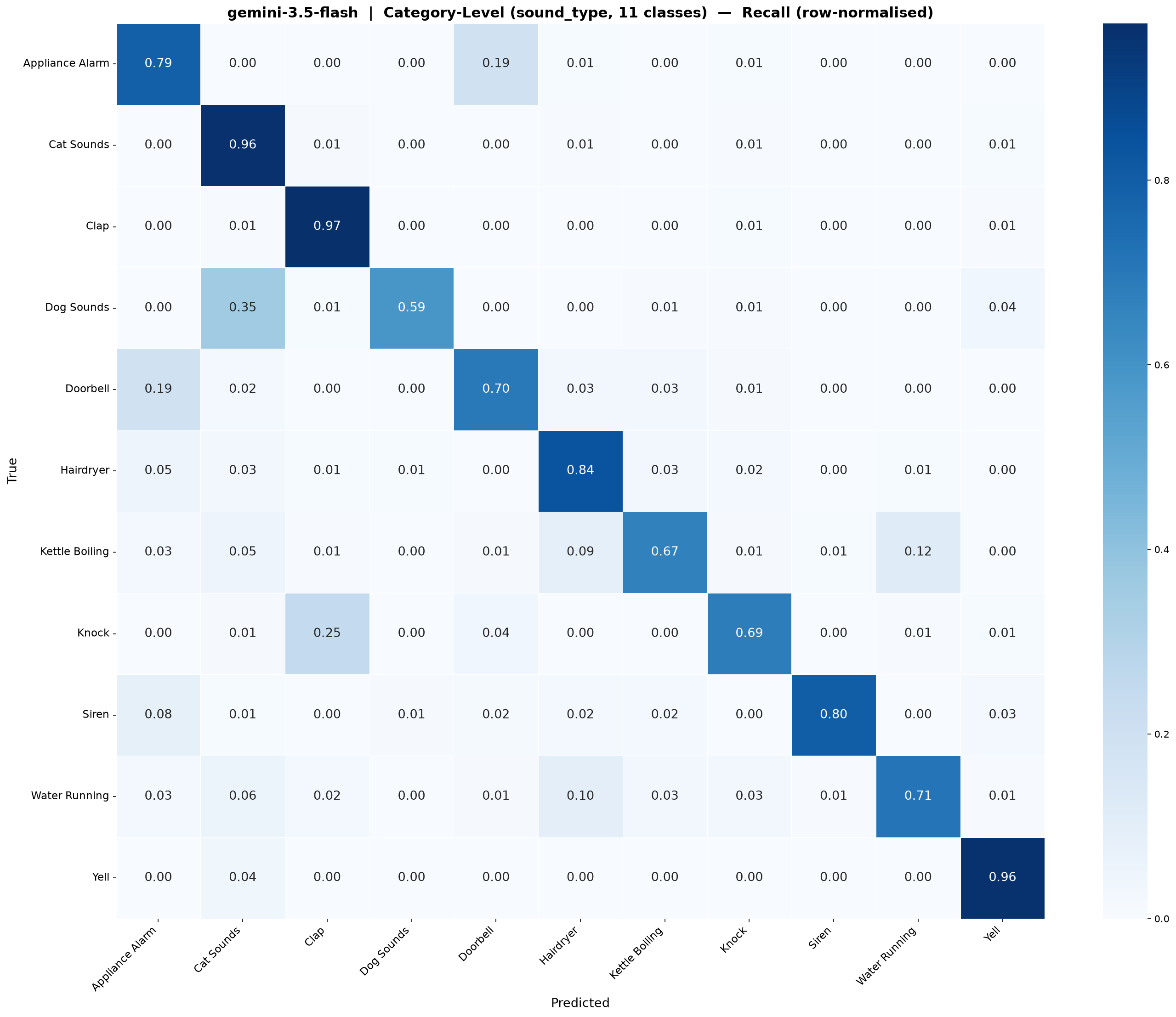}
  \caption{gemini-3.5-flash --- category-level confusion matrix (recall).}
\end{figure}
\begin{figure}[H]
  \centering
  \includegraphics[width=0.85\textwidth]{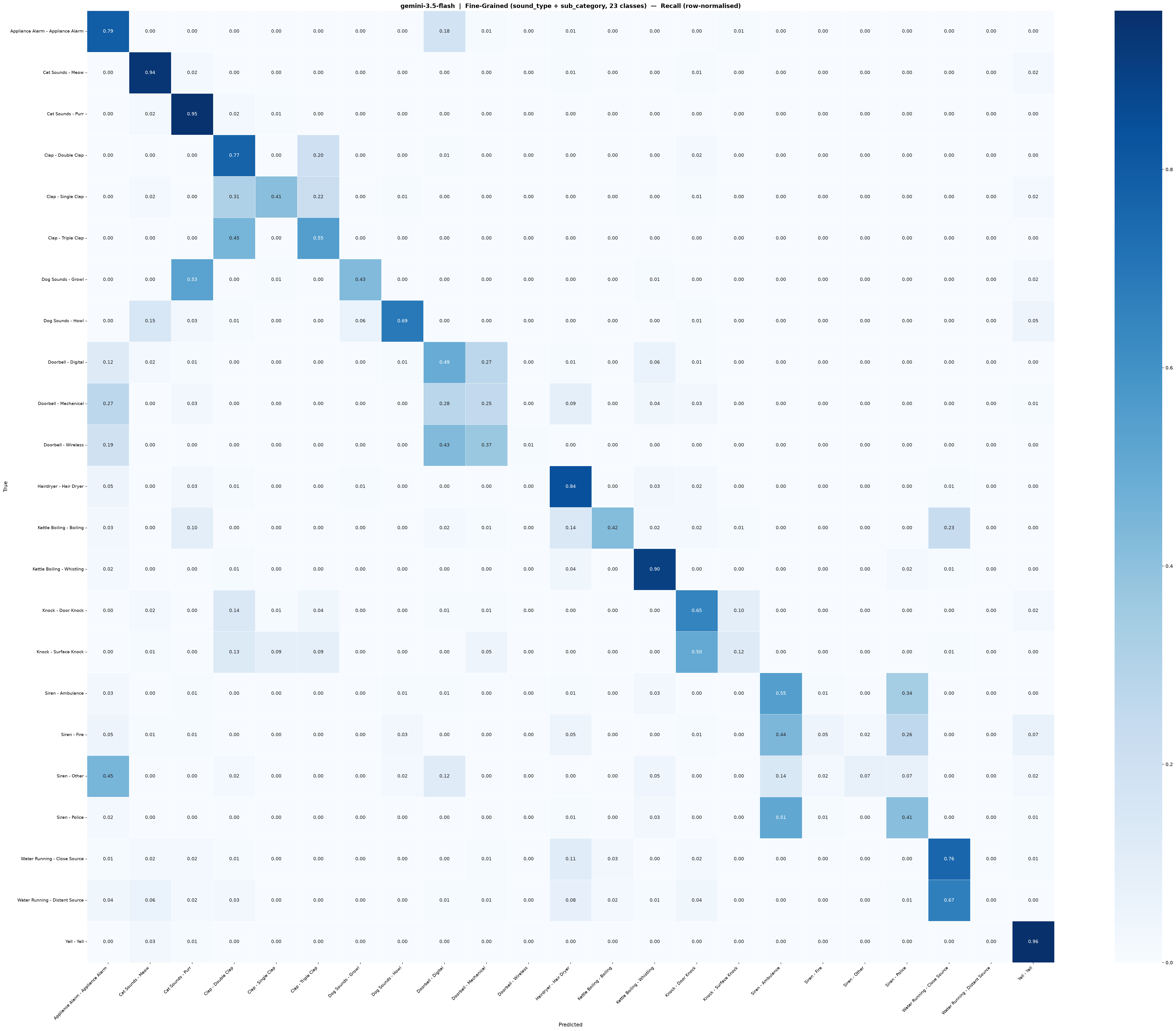}
  \caption{gemini-3.5-flash --- fine-grained confusion matrix (recall).}
\end{figure}

\begin{figure}[H]
  \centering
  \includegraphics[width=0.8\textwidth]{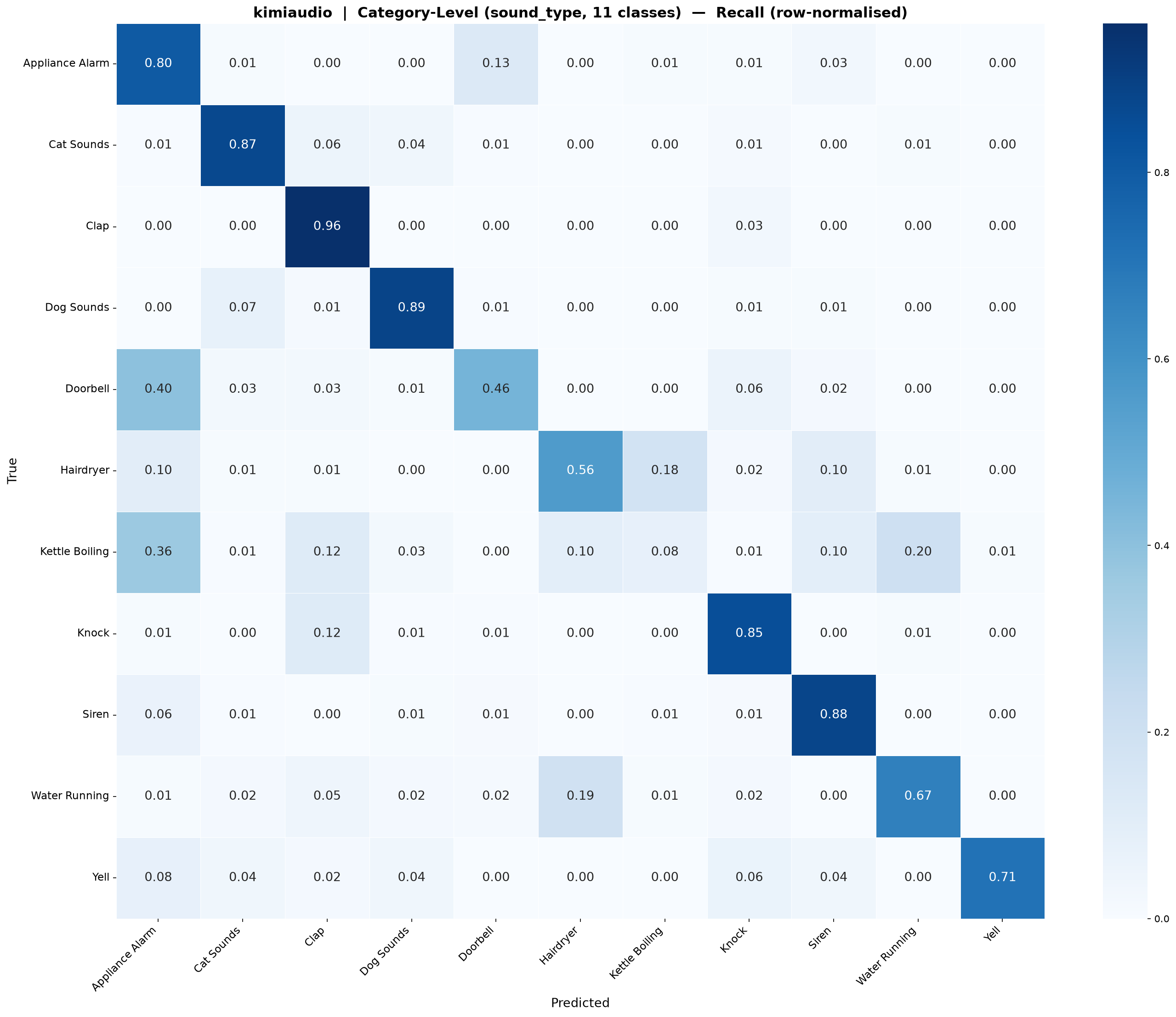}
  \caption{kimi-audio-7b-instruct (Tier A) --- category-level confusion matrix
  (recall). $N{=}2{,}206$; 36 samples with no scoreable answer excluded (see
  Section~\ref{sec:methodology}).}
\end{figure}
\begin{figure}[H]
  \centering
  \includegraphics[width=0.85\textwidth]{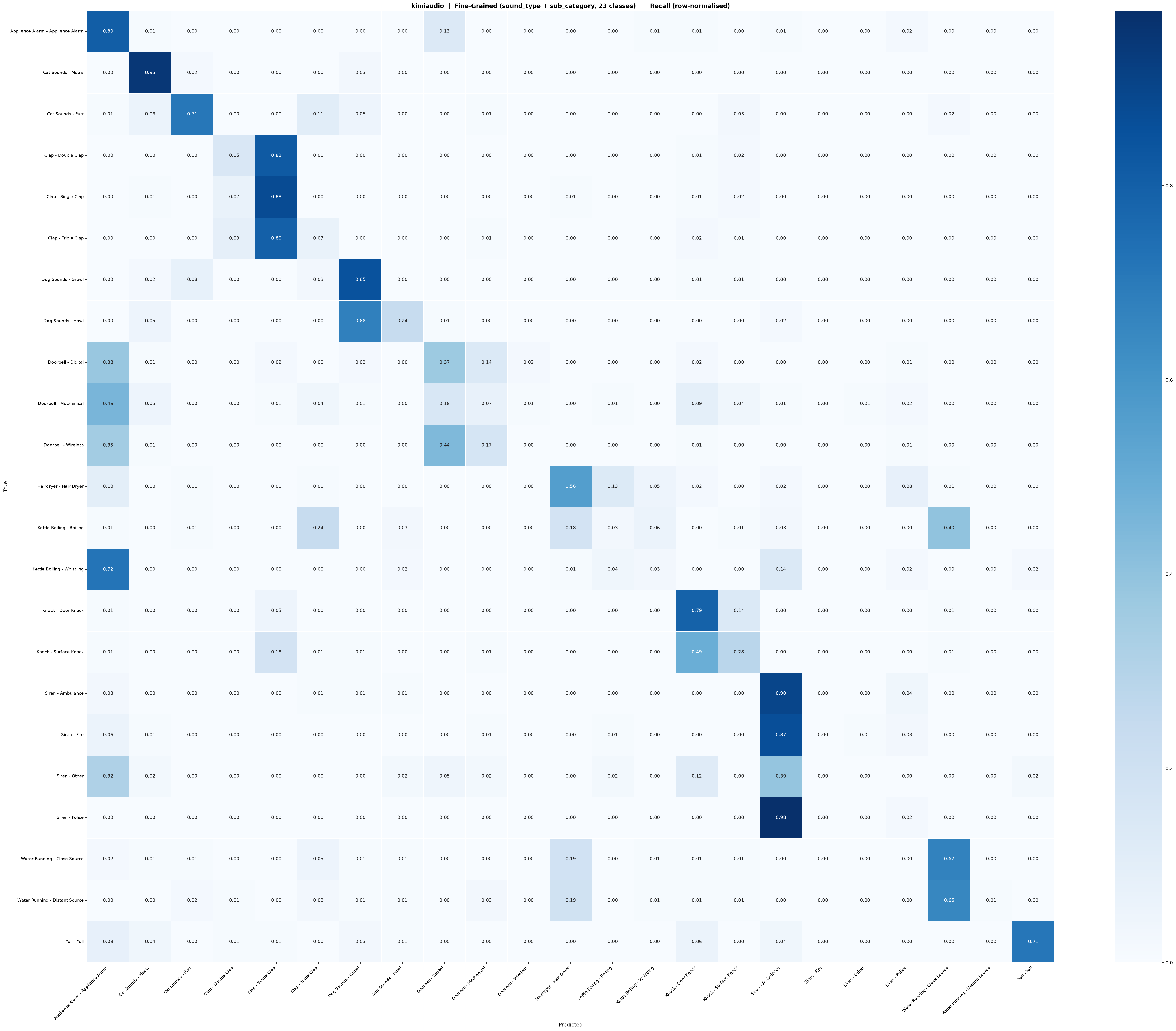}
  \caption{kimi-audio-7b-instruct (Tier A) --- fine-grained confusion matrix
  (recall). $N{=}2{,}206$.}
\end{figure}

\begin{figure}[H]
  \centering
  \includegraphics[width=0.8\textwidth]{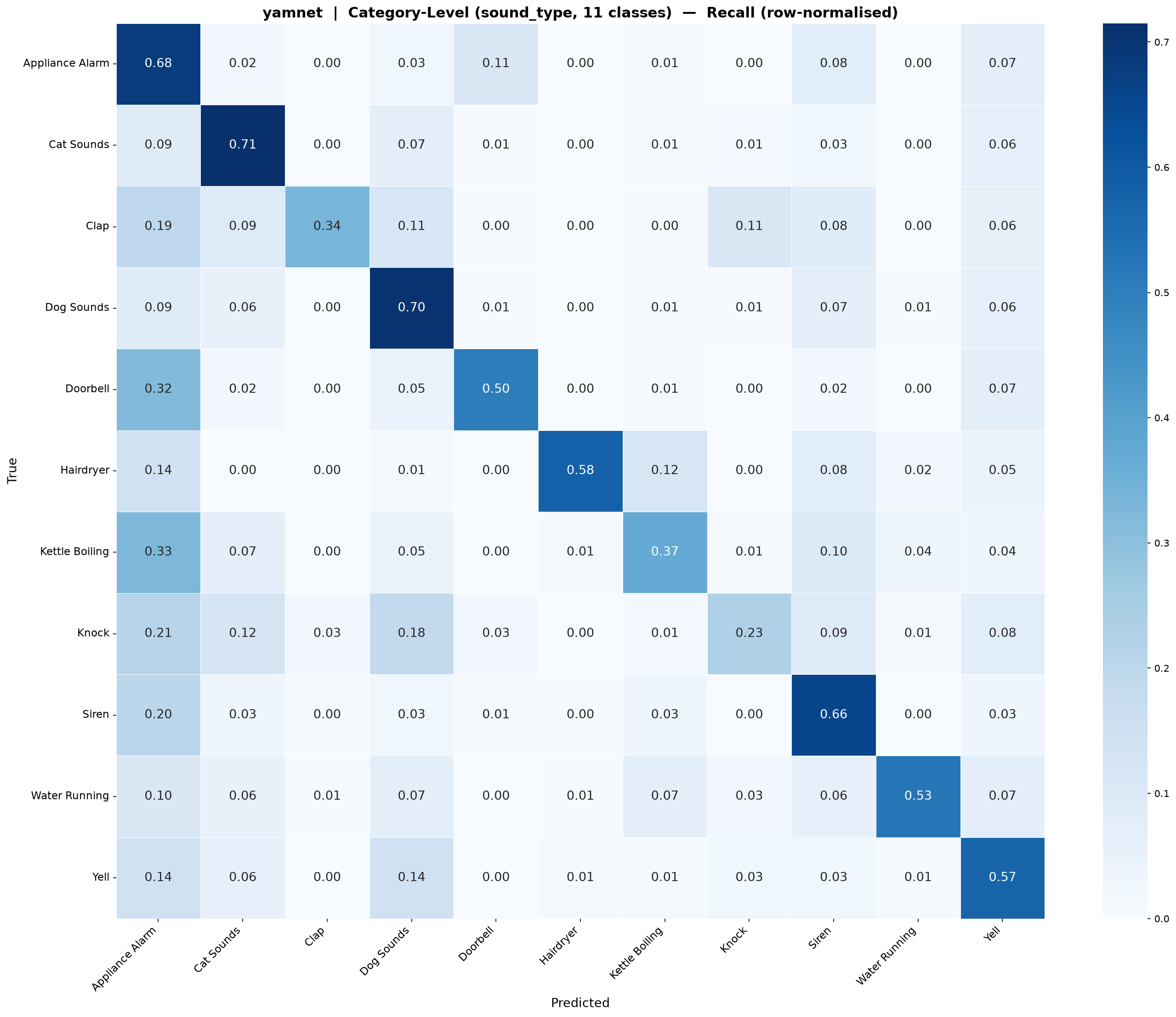}
  \caption{YAMNet (Tier B) --- category-level confusion matrix (recall).}
\end{figure}
\begin{figure}[H]
  \centering
  \includegraphics[width=0.85\textwidth]{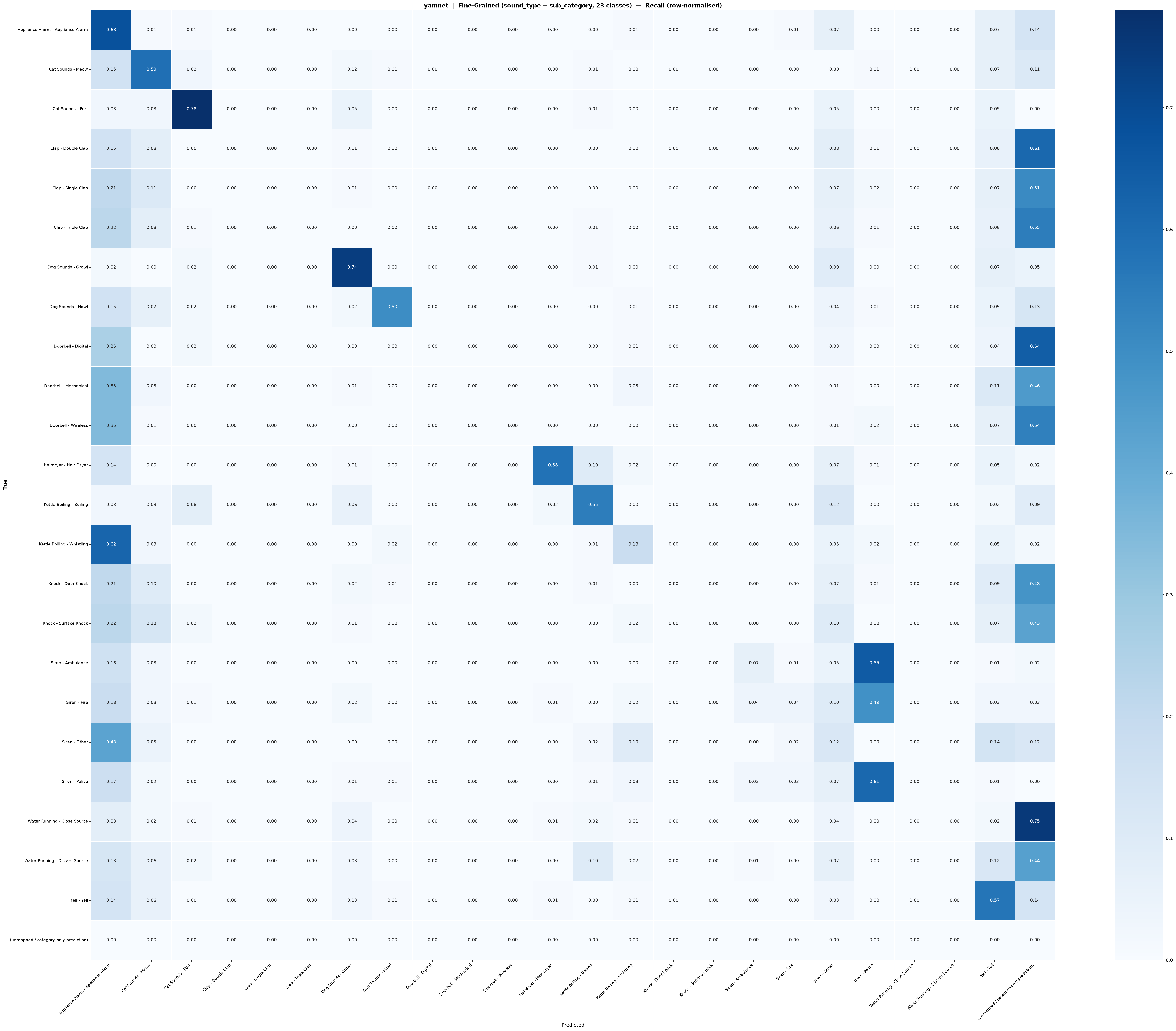}
  \caption{YAMNet (Tier B) --- fine-grained confusion matrix (recall), Top-1.}
\end{figure}
\begin{figure}[H]
  \centering
  \includegraphics[width=0.85\textwidth]{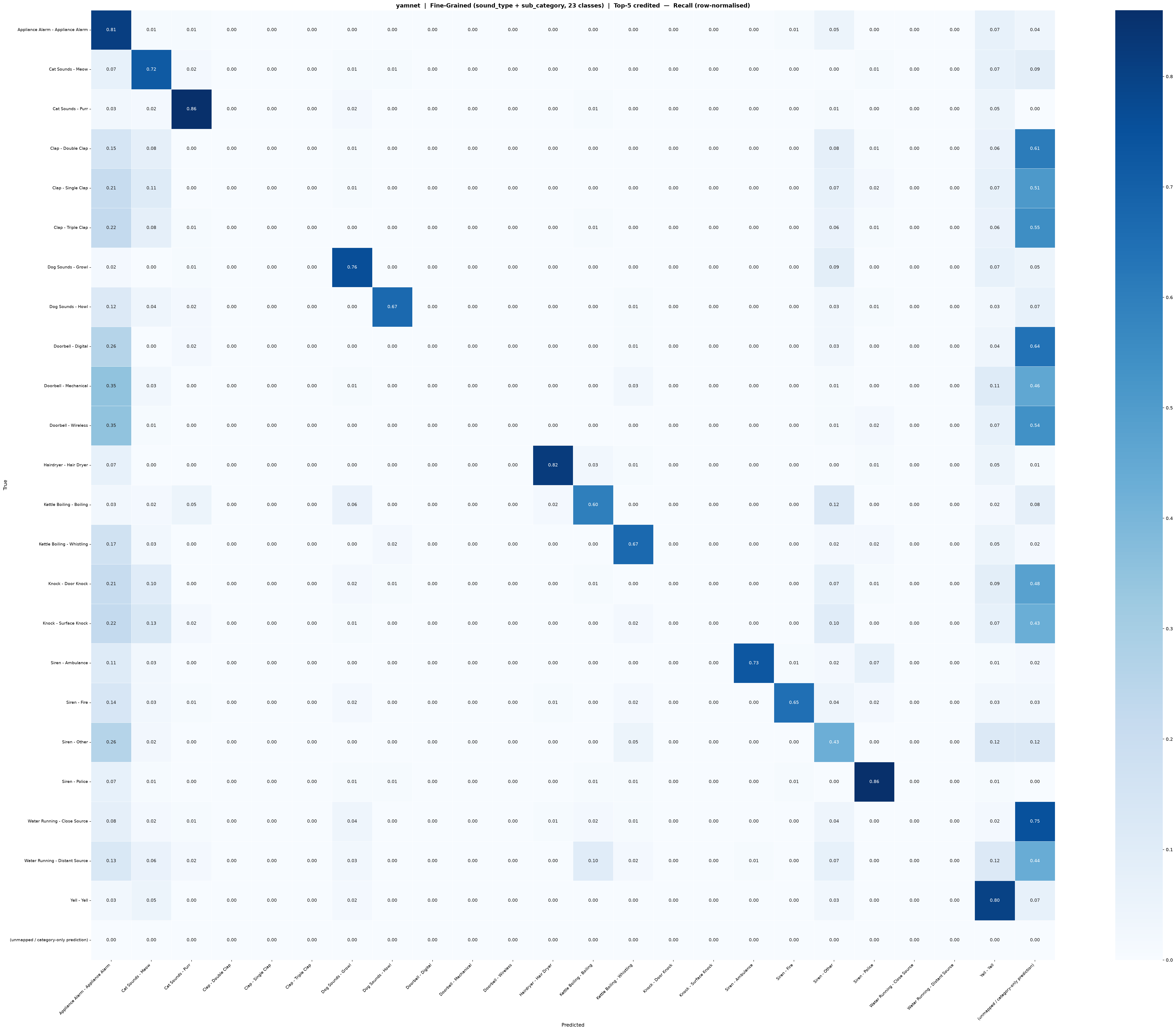}
  \caption{YAMNet (Tier B) --- fine-grained confusion matrix (recall), \textbf{Top-5
  credited}: diagonal counts a sample as correct if the true label appears anywhere
  in the model's top-5 ranked mapped predictions, otherwise its actual Top-1 guess
  is shown (see Section~\ref{sec:topk}).}
\end{figure}

\begin{figure}[H]
  \centering
  \includegraphics[width=0.8\textwidth]{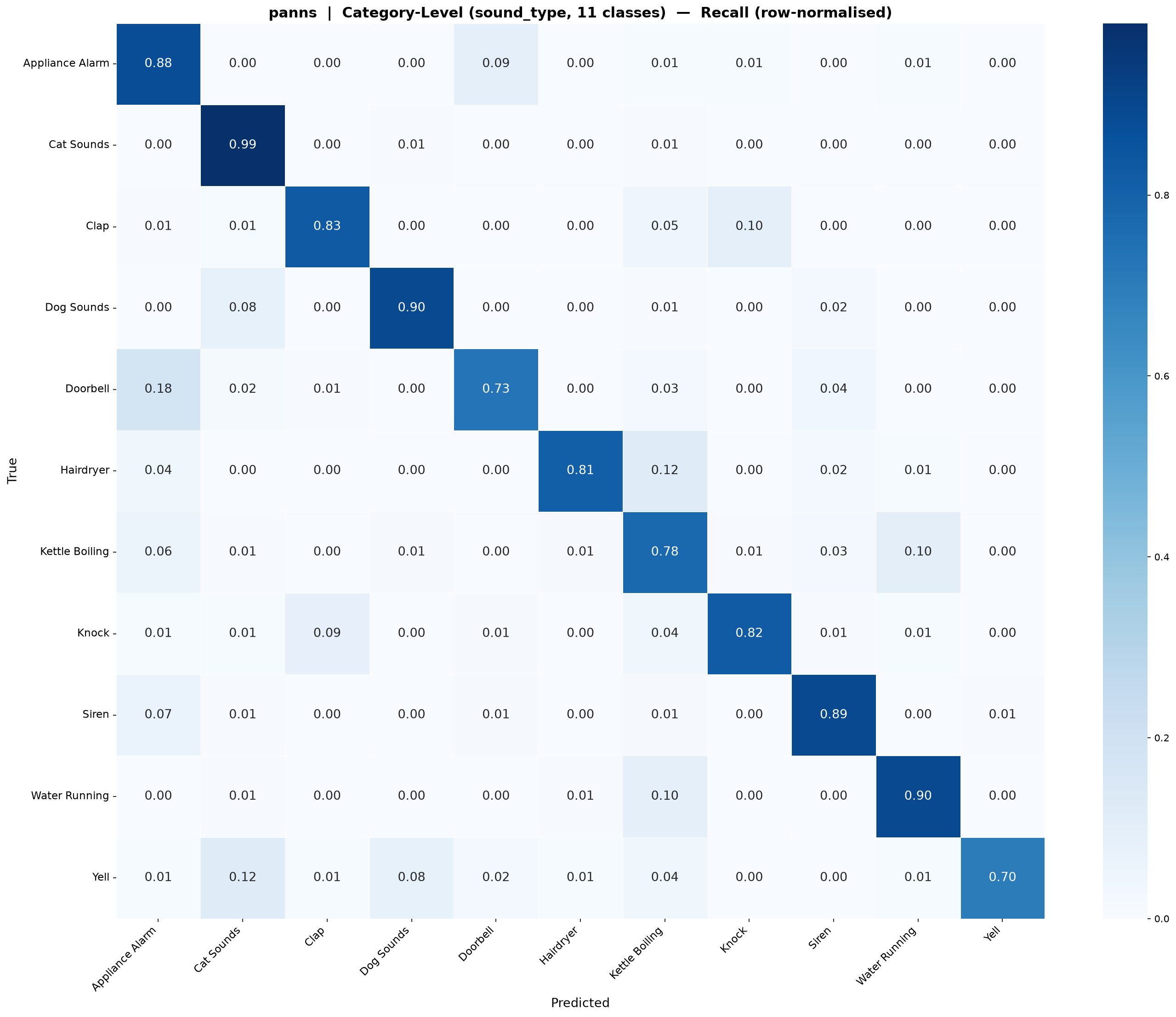}
  \caption{PANNs (Tier B) --- category-level confusion matrix (recall).}
\end{figure}
\begin{figure}[H]
  \centering
  \includegraphics[width=0.85\textwidth]{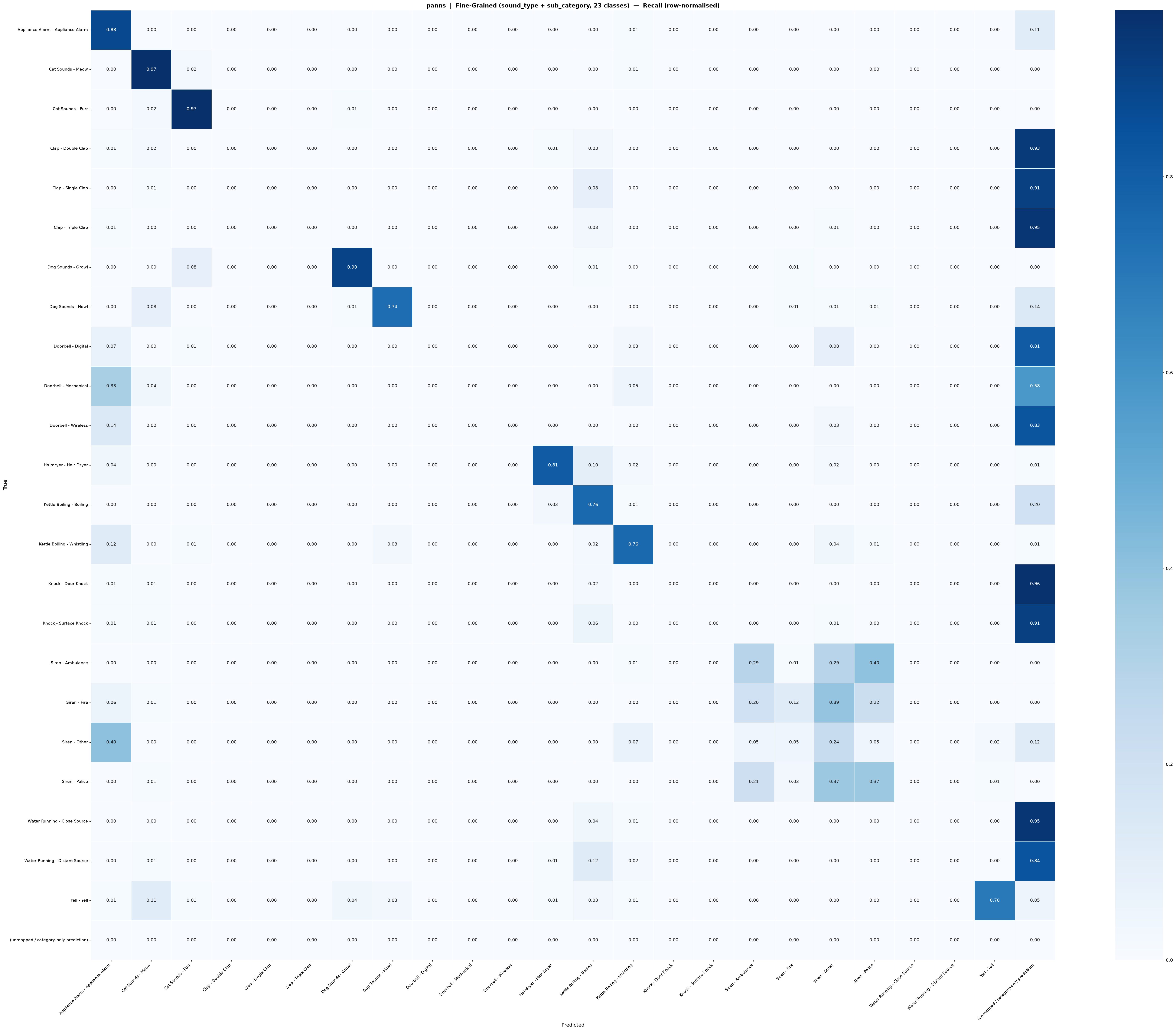}
  \caption{PANNs (Tier B) --- fine-grained confusion matrix (recall), Top-1.}
\end{figure}
\begin{figure}[H]
  \centering
  \includegraphics[width=0.85\textwidth]{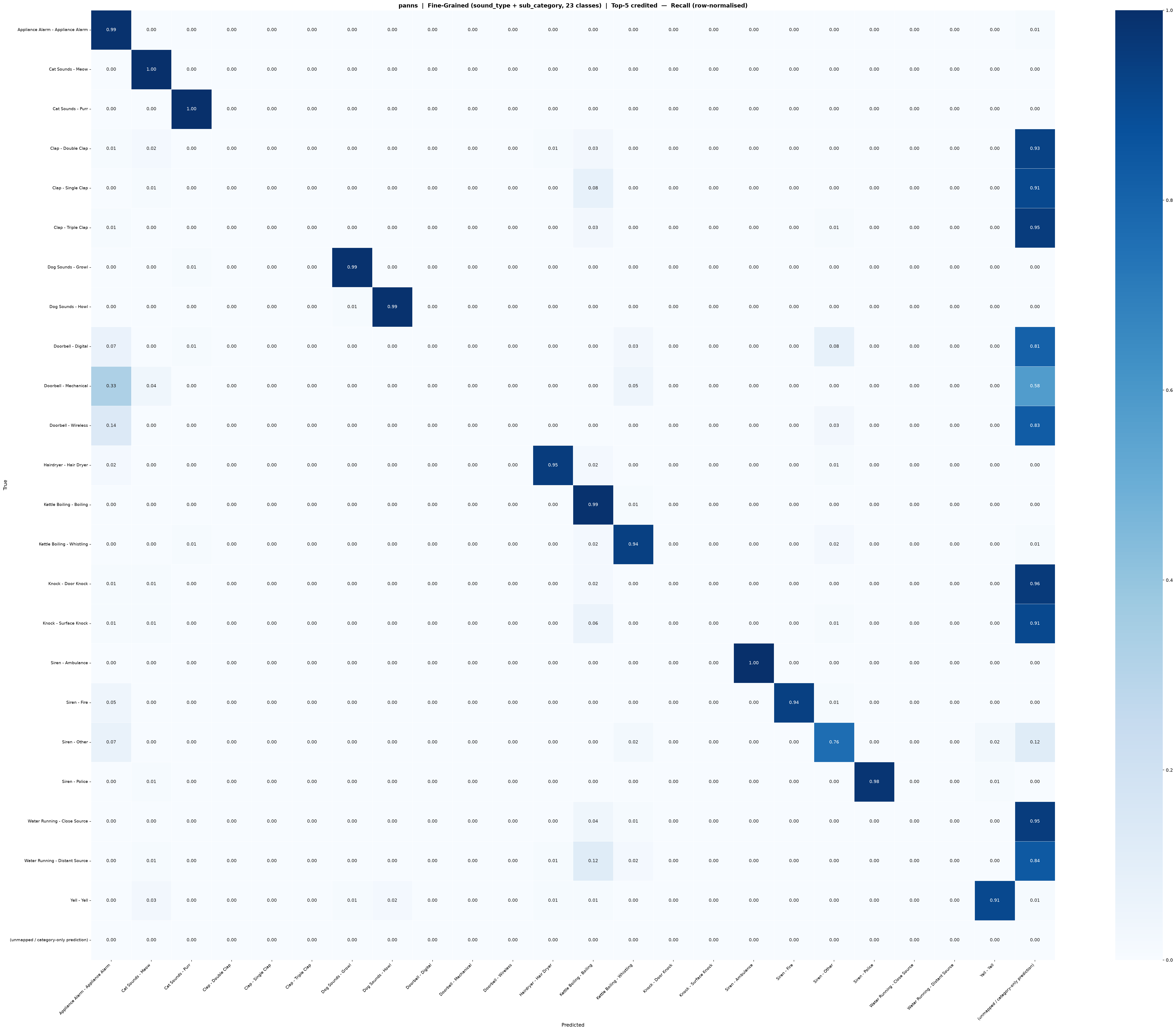}
  \caption{PANNs (Tier B) --- fine-grained confusion matrix (recall), \textbf{Top-5
  credited} (see Section~\ref{sec:topk}).}
\end{figure}

\begin{figure}[H]
  \centering
  \includegraphics[width=0.8\textwidth]{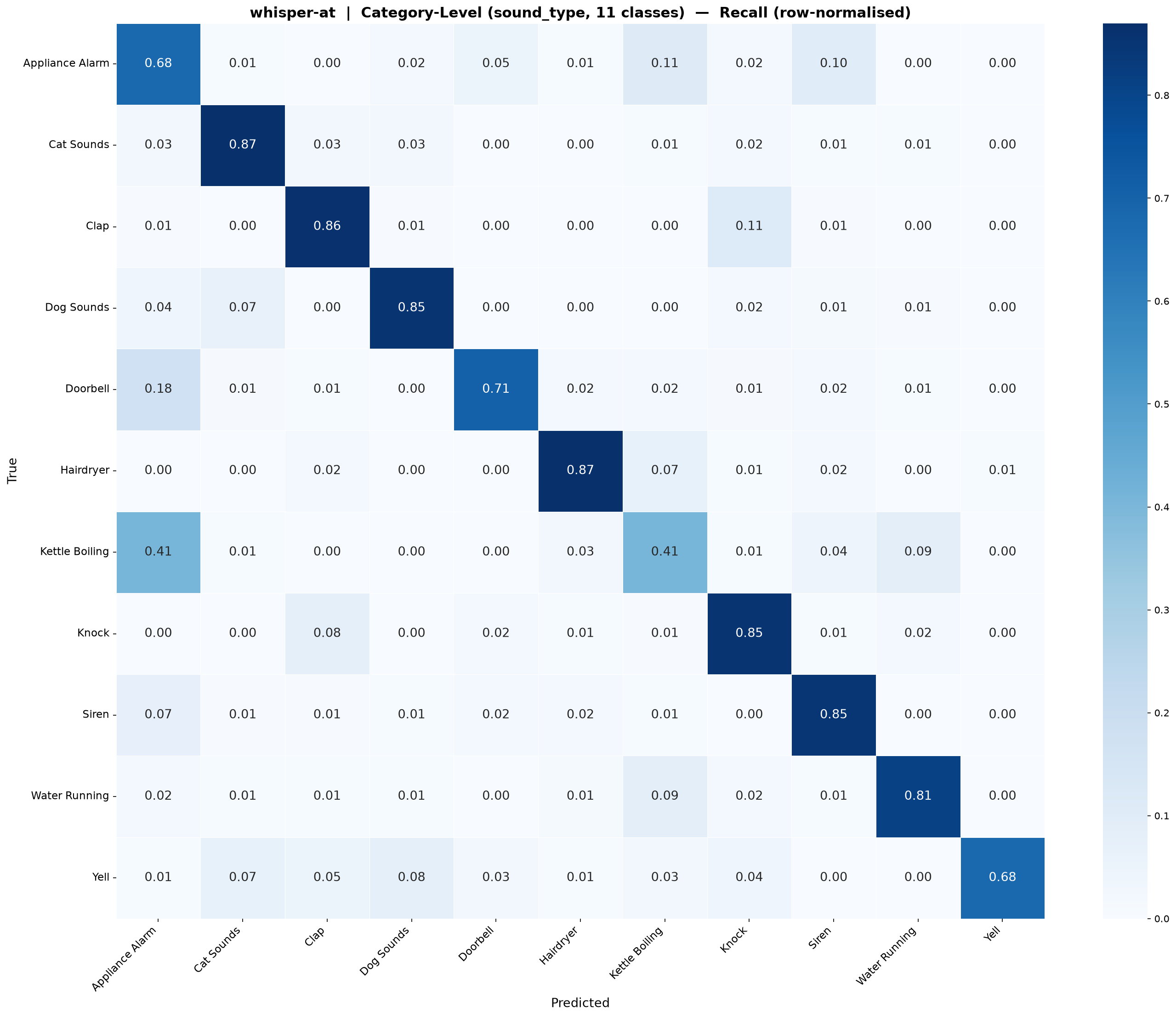}
  \caption{Whisper-AT (Tier B) --- category-level confusion matrix (recall).}
\end{figure}
\begin{figure}[H]
  \centering
  \includegraphics[width=0.85\textwidth]{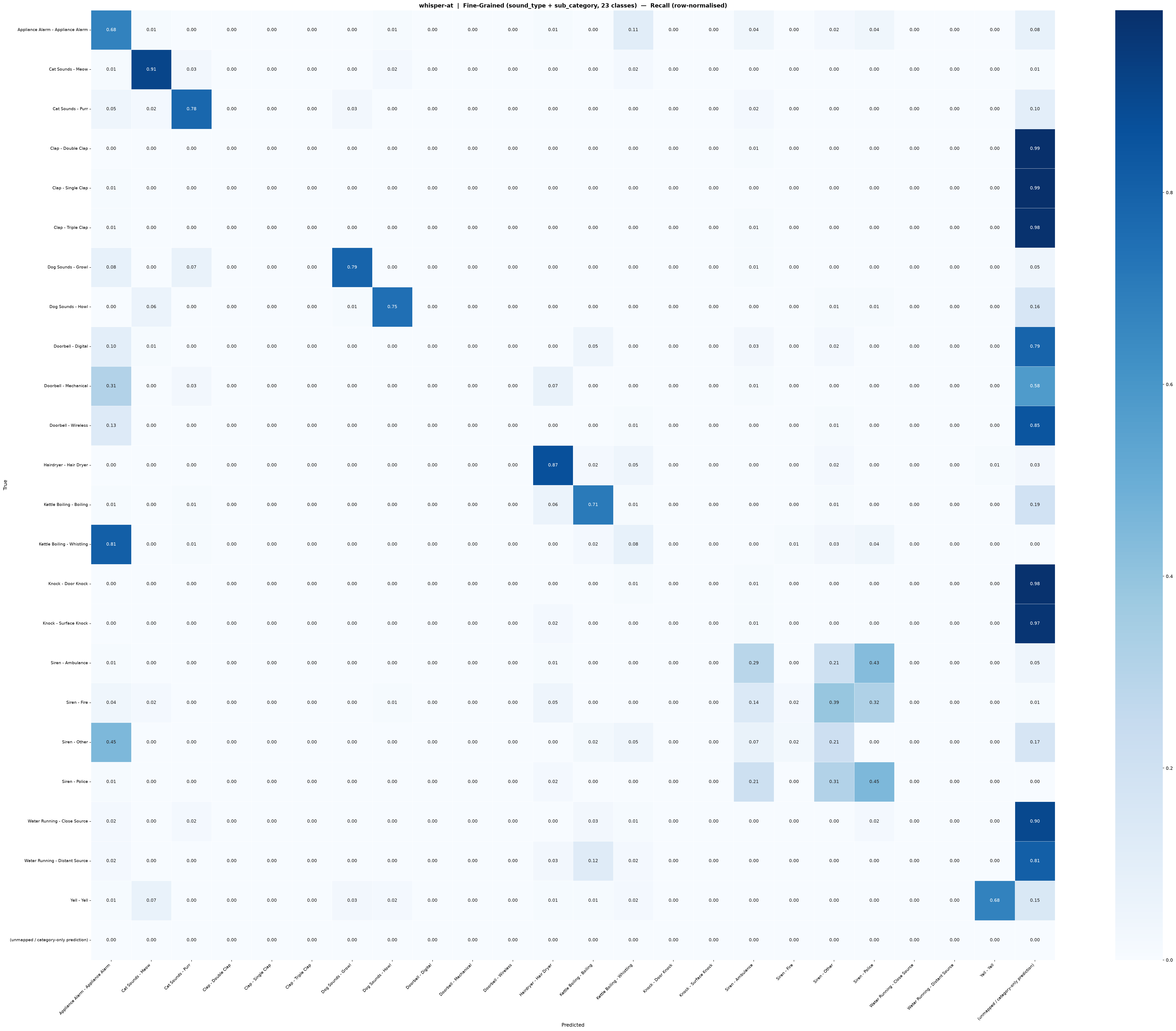}
  \caption{Whisper-AT (Tier B) --- fine-grained confusion matrix (recall), Top-1.}
\end{figure}
\begin{figure}[H]
  \centering
  \includegraphics[width=0.85\textwidth]{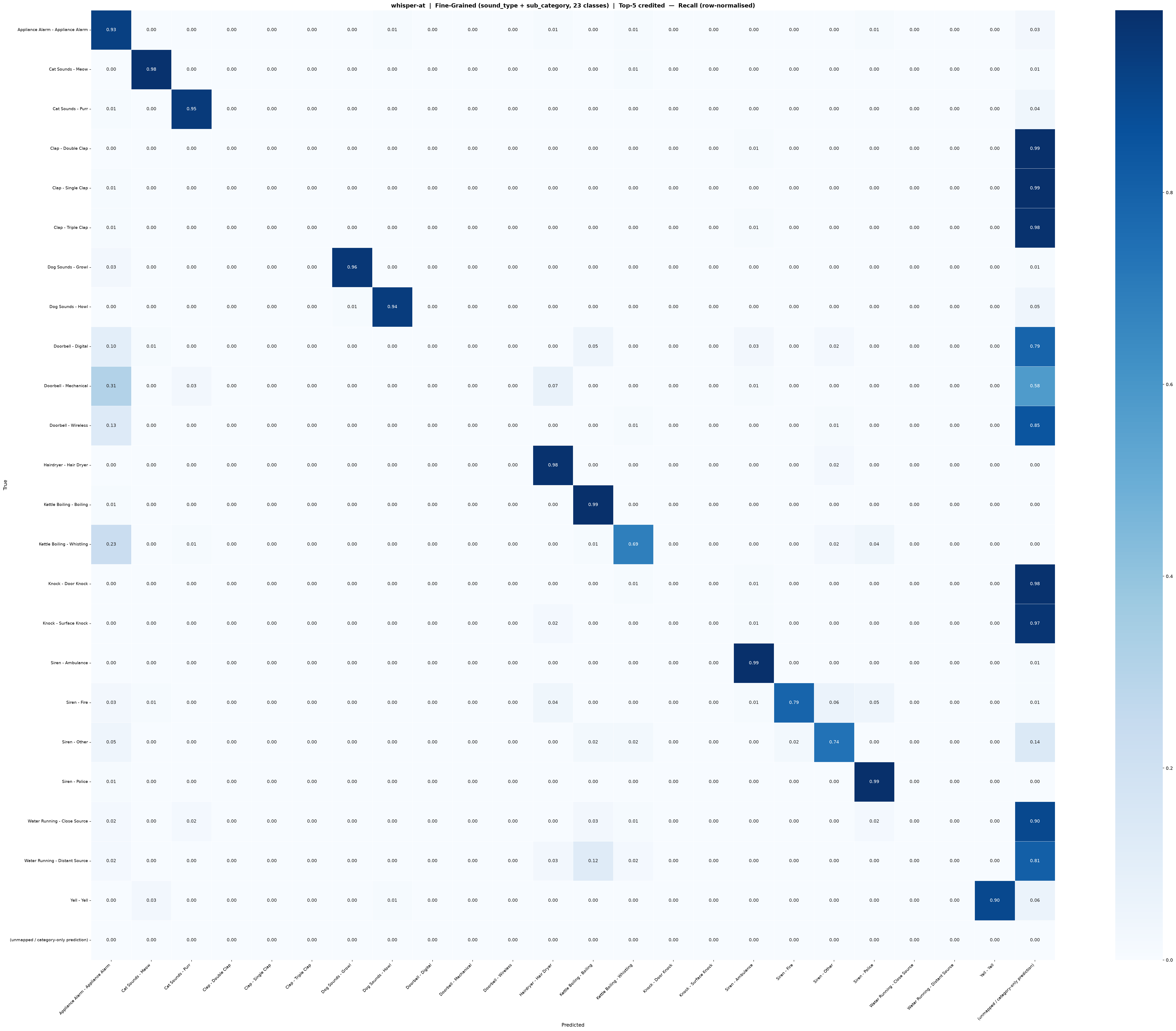}
  \caption{Whisper-AT (Tier B) --- fine-grained confusion matrix (recall),
  \textbf{Top-5 credited} (see Section~\ref{sec:topk}).}
\end{figure}

\begin{figure}[H]
  \centering
  \includegraphics[width=0.8\textwidth]{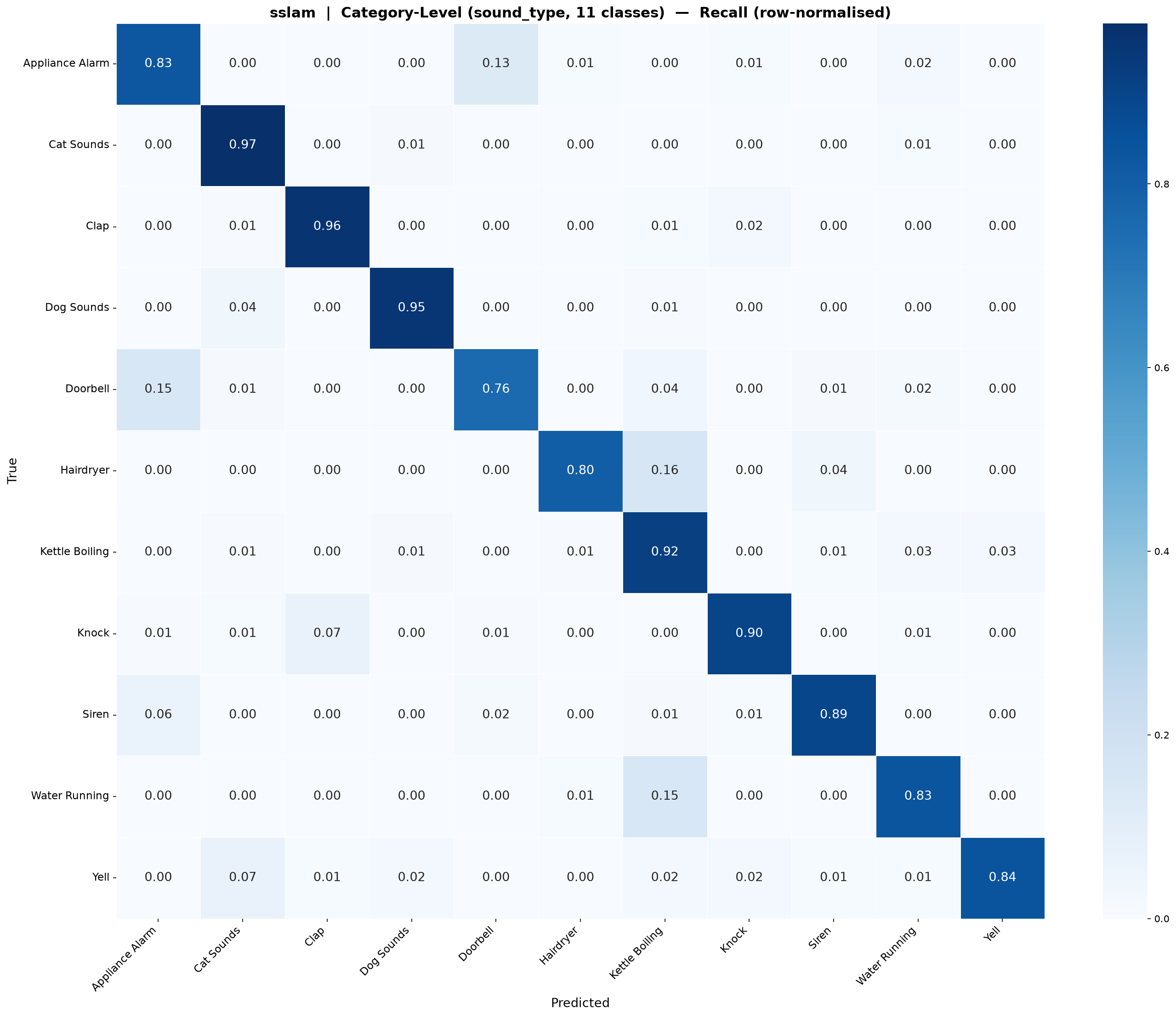}
  \caption{SSLAM (Tier B) --- category-level confusion matrix (recall).}
\end{figure}
\begin{figure}[H]
  \centering
  \includegraphics[width=0.85\textwidth]{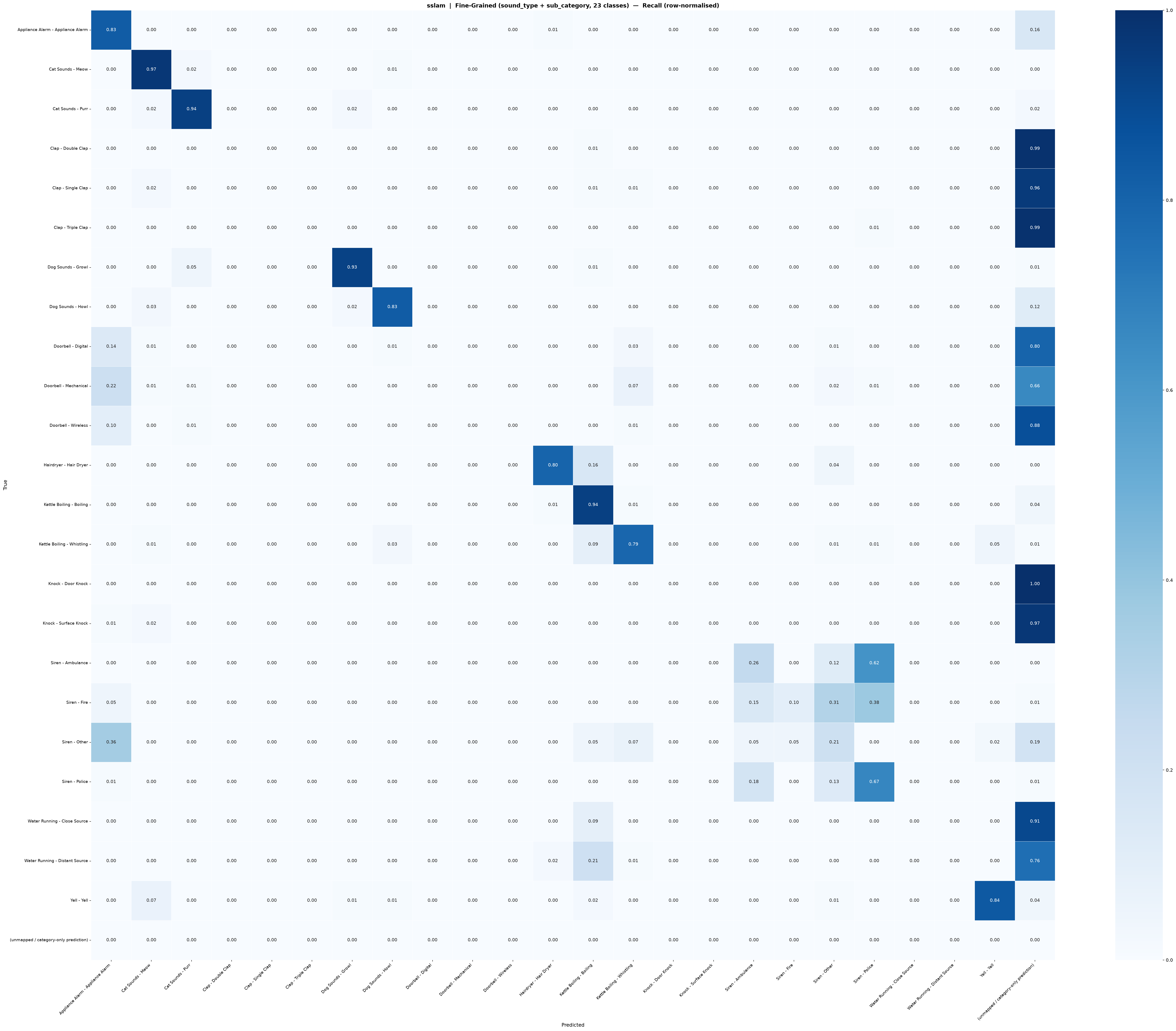}
  \caption{SSLAM (Tier B) --- fine-grained confusion matrix (recall), Top-1.}
\end{figure}
\begin{figure}[H]
  \centering
  \includegraphics[width=0.85\textwidth]{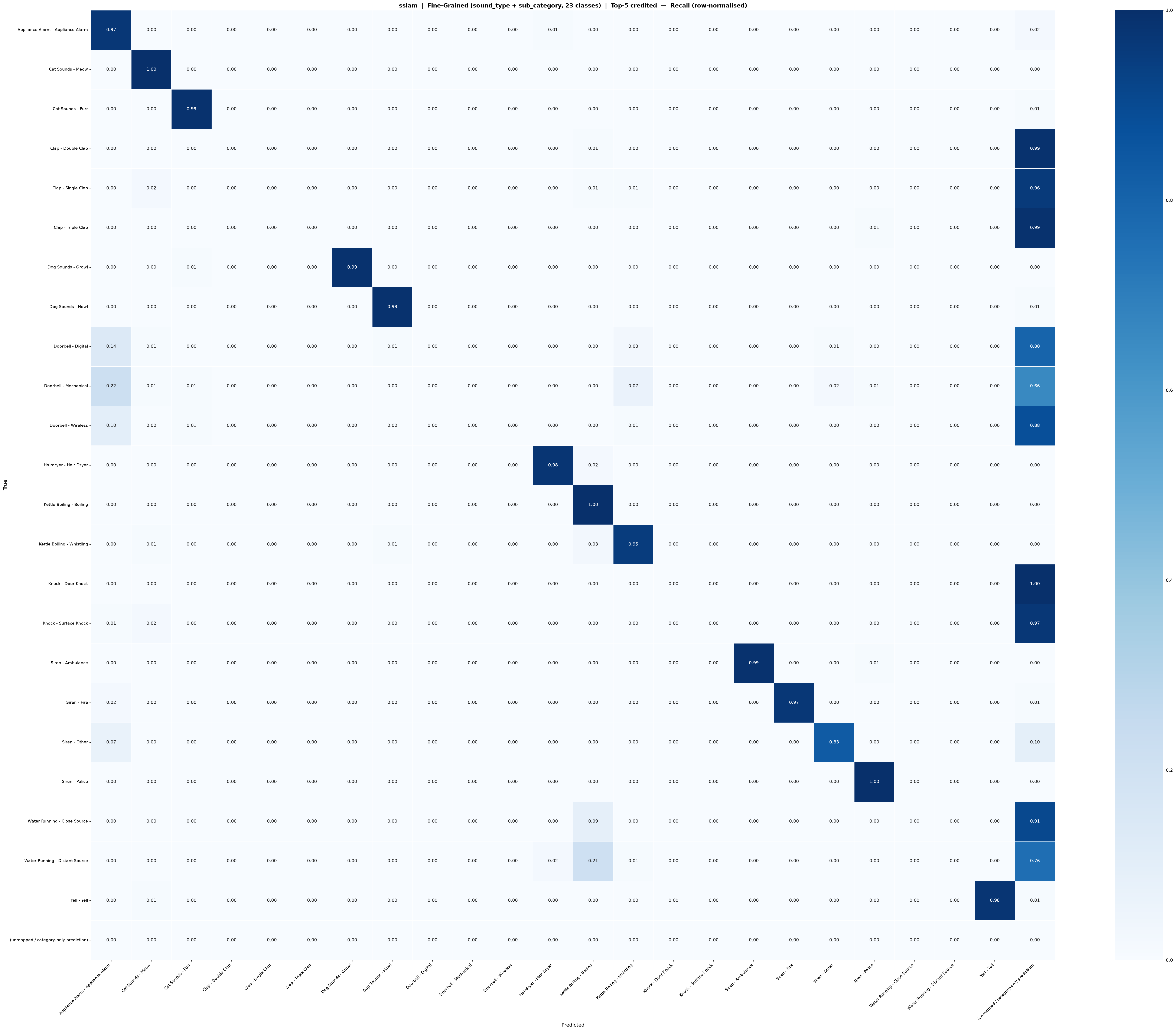}
  \caption{SSLAM (Tier B) --- fine-grained confusion matrix (recall), \textbf{Top-5
  credited} (see Section~\ref{sec:topk}).}
\end{figure}

\begin{figure}[H]
  \centering
  \includegraphics[width=0.8\textwidth]{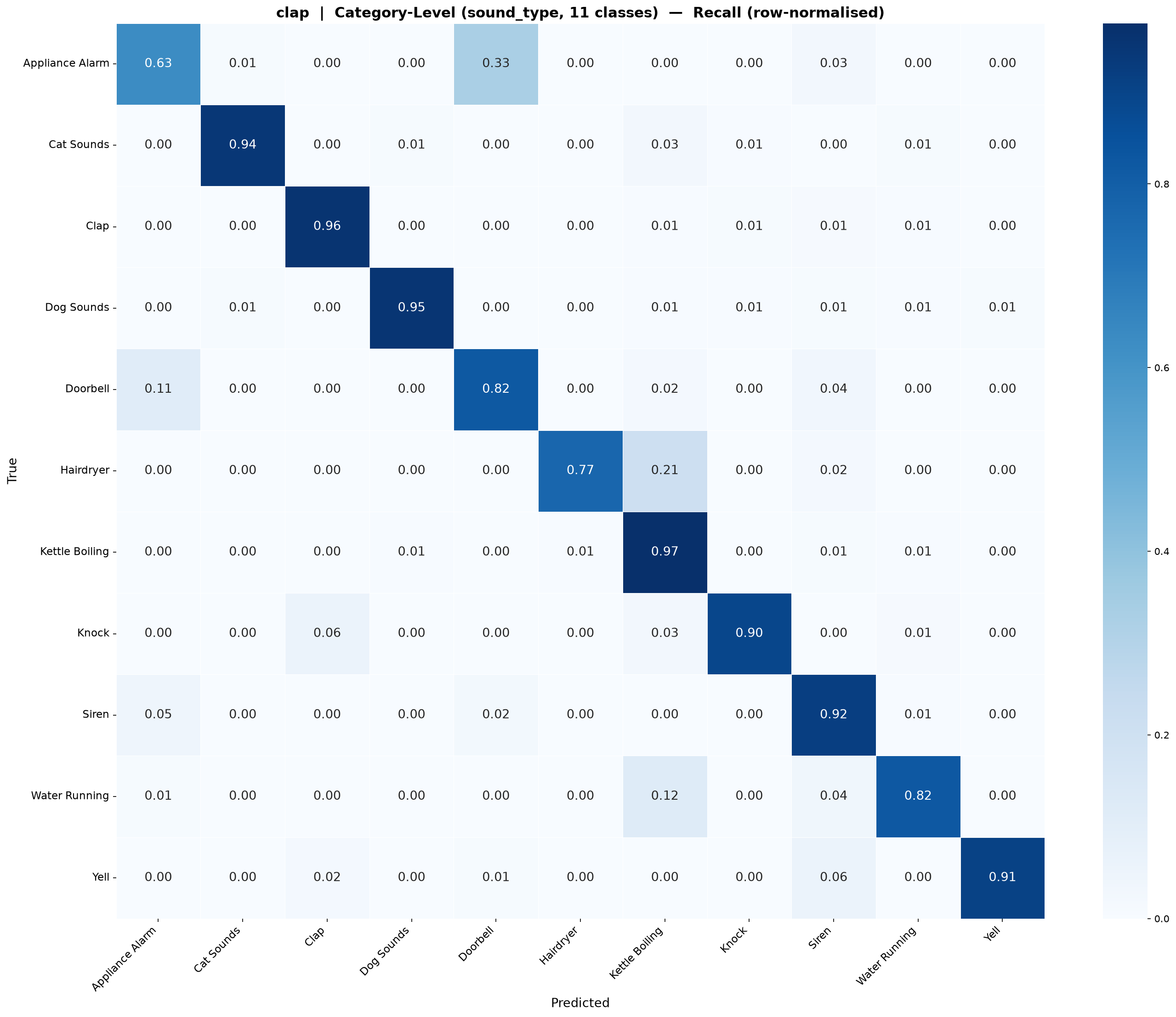}
  \caption{CLAP (Tier C) --- category-level confusion matrix (recall).}
\end{figure}
\begin{figure}[H]
  \centering
  \includegraphics[width=0.85\textwidth]{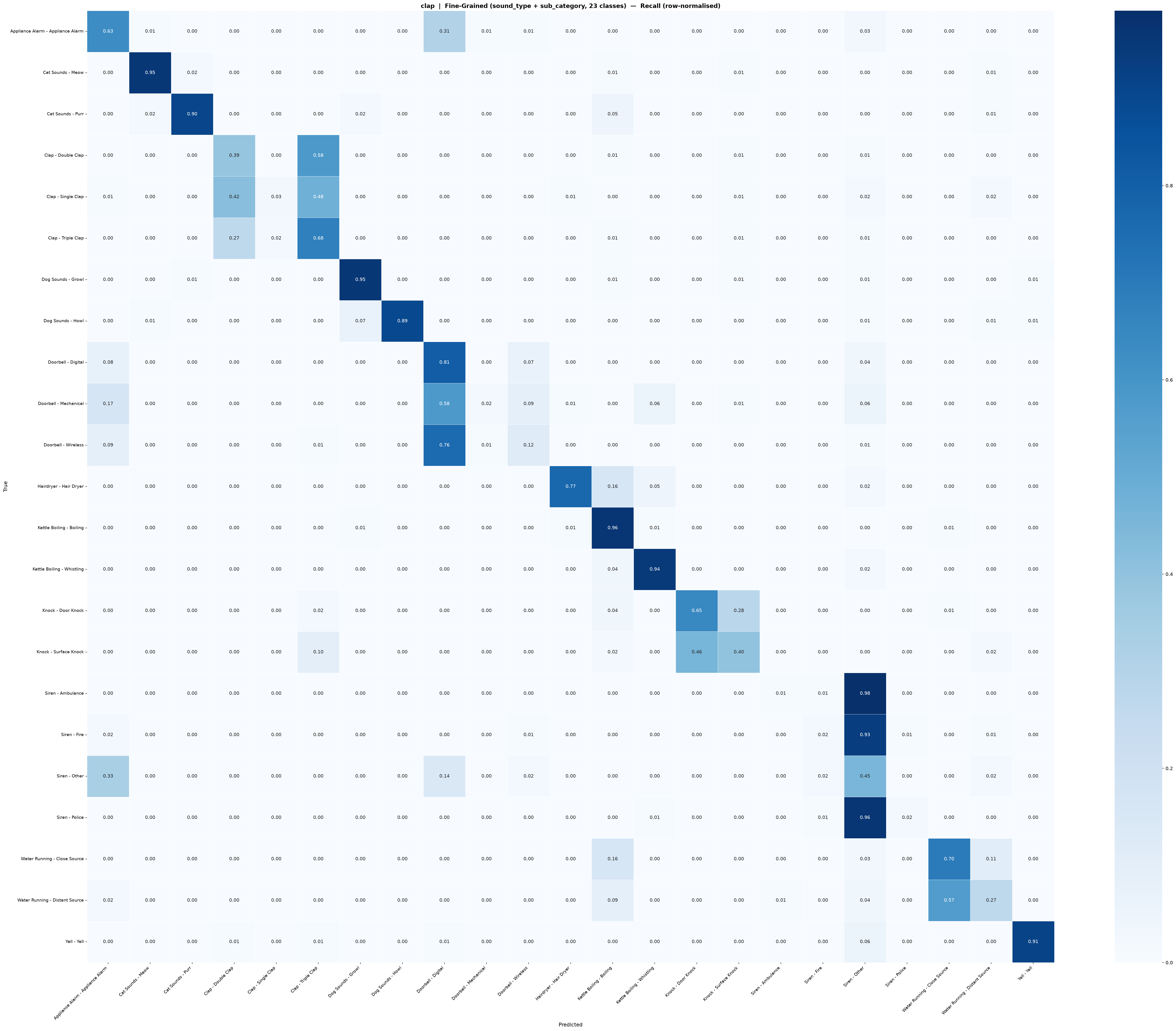}
  \caption{CLAP (Tier C) --- fine-grained confusion matrix (recall).}
  \label{fig:cm-a20}
\end{figure}

\end{document}